\documentclass[preprint]{aastex}

\slugcomment{Submitted to the Astrophysical Journal}
\shorttitle{Micro-lensing calibration of stellar masses}
\shortauthors{Schechter et al.}
\begin{document}
\title{A calibration of the stellar mass fundamental plane at z $\sim$ 0.5
using the micro-lensing induced flux ratio anomalies of
macro-lensed quasars\altaffilmark{*,\S,\S\S}}

\author{Paul L. Schechter\altaffilmark{a,b}}
\affil{MIT Kavli Institute for Astrophysics and Space Research,
Cambridge, MA 02139}

\author{David Pooley\altaffilmark{c}}
\affil{Sam Houston State University, Huntsville, TX 77341}

\author{Jeffrey A. Blackburne}
\affil{Aret\'e Associates, Northridge, CA 91324}

\author{Joachim Wambsganss\altaffilmark{d}}
\affil{Zentrum f\"ur Astronomie der Universit\"at Heidelberg} 

\altaffiltext{*}
{The scientific results reported in this article are based
to a significant degree on observations made with the Chandra X-ray Observatory
and published previously in cited articles.}
\altaffiltext{\S}
{This paper includes data gathered with the 6.5 meter Magellan 
telescopes located at Las Campanas Observatory, Chile.}
\altaffiltext{\S\S}{Based in part on observations made with the NASA/ESA 
Hubble Space Telescope, obtained at the Space Telescope Science Institute,
which is operated by the Association of Universities for Research in
Astronomy, Inc., under NASA contract NAS5-26555.  These observations
are associated with program GO-9854.}
\altaffiltext{a}{MIT Department of Physics}
\altaffiltext{b}{Williams College Deparment of Astronomy}
\altaffiltext{c}{Eureka Scientific}
\altaffiltext{d}{Pauli Center for Theoretical Physics, Z\"urich, Switzerland}

\begin{abstract}
We measure the stellar mass surface densities of early
type galaxies by observing the micro-lensing of 
macro-lensed quasars caused by individual stars,
including stellar remnants, brown dwarfs and 
red dwarfs too faint to produce photometric or 
spectroscopic signatures.  Instead of observing
multiple micro-lensing events in a single system, we 
combine single epoch X-ray snapshots of ten quadruple systems,
and compare the measured relative magnifications
for the images with those computed from macro-models.
We use these to  normalize a stellar mass  
fundamental plane constructed using a Salpeter
IMF with a low mass cutoff of $0.1 M_\odot$ 
and treat the zeropoint of the surface mass
density as a free parameter.  Our method measures
the {\it graininess} of the gravitational potential
produced by individual stars, in contrast to
methods that decompose a smooth total
gravitational potential into two smooth 
components, one stellar and one dark.  We find
the median likelihood value for the normalization factor $\cal F$ by which
the Salpeter stellar masses must be multiplied is 1.23,
with a one sigma confidence range, dominated
by small number statistics, of $0.77 <{\cal F}< 2.10$.
\end{abstract}

\section {Introduction}

\subsection{Stellar masses from micro-lensing}

So-called ``stellar" masses for early type galaxies are almost always
determined by one of two methods: either they are estimated from
spectra (sometimes only broad band colors) 
or they are deduced by
subtracting the contribution of an assumed dark matter component from
a combined mass inferred from kinematic (and sometimes macro-lensing)
measurements.  The two methods, with their many variants,
are described at length in the review by Courteau et al. (2014).

Both methods have shortcomings.  Spectral methods suffer from the fact
that lower main sequence stars, substellar objects and stellar remnants
contribute negligibly to the observed light, and therefore cannot be
detected in spectra.  Such determinations must therefore rely on some
assumed shape for the stellar mass function.  To quote from a 
frequently cited example of these efforts (Kauffmann et al. 2003)

{\parindent=20pt {\narrower ``All of our derived parameters are tied
    to a specific choice of IMF [Initial Mass Function].  Changing the
    IMF would scale the stellar mass estimates by a fixed factor.  For
    example, changing from a Kroupa (2001) to a Salpeter IMF with a
    cutoff at $0.1 M_\odot$ would result in a factor of 2 increase in
    the stellar mass."

\par
}
}
Dark matter subtraction techniques are vulnerable to mistaken
assumptions about the density profile of the dark matter and to
mistaken assumptions about the constancy or non-constancy of the
mass-to-light ratio of the stellar component.  A recent paper by
Cappellari et al. (2012) illustrates the effects of varying assumptions
about the shapes of dark matter halos.

In this paper we use a third method: determining the stellar mass
surface density of an early type galaxy from brightness fluctuations
of the four images of a background quasar that is 
both multiply-imaged (``macro-lensed'') by the galaxy and 
micro-lensed by the individual stars in that galaxy (Schechter and Wambsganss
2004; Kochanek 2004).  This method, in contrast to spectral methods,
is sensitive to stellar mass near and below the hydrogen burning
limit, as well as to mass in stellar remnants.  And where dark matter
subtraction methods make asumptions about the dark matter profiles,
the gravitational micro-lensing technique makes only an assumption
about the combined gravitational potential, one that has been
subjected to extensive observational verification.

Micro-lensing produces flux ratio anomalies of the sort described by
Schechter and Wambsganss (2002).  Ideally one would observe a single
system long enough to see a great many fluctuations and infer an
accurate stellar surface density.  But the timescale for micro-lensing
variations is of order ten years for a lens at redshift $z \sim 0.5$
(Mosquera and Kochanek 2011), and observations of even four quasar
images at a single epoch give only broad constraints on the mass
surface density via the deviation of the observed flux ratios from those
predicted by a macro-model.  So instead we observe a number of
systems at a single epoch and combine results.  

Such a measurement is no different, in principle, from the measurement
of the density of massive compact halo objects in the Milky way via
micro-lensing measurements of stars in the Magellanic Clouds
(Paczy\'nski 1986b), though it differs in the nature of the source that
is lensed and in the very much higher optical depth (Paczy\'nski 1986a;
Witt, Mao and Schechter 1995) along the lines of sight to multiply
imaged quasars.

There is as yet no theory that predicts in closed form the statistics
of such light curves as a function of point mass density and external
shear, at least not at the optical depths we consider.  One must
therefore carry out simulations for each case of interest, using the
ray shooting technique pioneered by Kayser, Refsdal and Stabell
(1986) and Schneider and Weiss (1987), and refined by Wambsganss (1990,
1999).

It is the peculiar property of such micro-lensing that the
instantaneous magnification probability distribution is determined
almost entirely by the surface mass density of micro-lenses, with a
dependence upon the distribution of masses that is so weak that only
with elaborately contrived simulations can it be observed at all
(Schechter, Lewis and Wambsganss 2004).

\subsection{The stellar mass fundamental plane}

The combination of a set of results from multiple quasars is non-trivial for
several reasons: a) the lens galaxies do not all have the same linear
sizes and stellar mass surface densities, b)
the quasar lines of sight do not all sample the same projected
distance from the lens, and c) the lenses lie at different redshifts.
Ideally our combined result would give the mass surface density for an early
type galaxy of a fiducial size at a fiducial radius and a fiducial
redshift.

The stellar mass fundamental plane (Hyde and Bernardi 2009)
can be used to scale measured stellar surface densities to a
common velocity dispersion (which we take to indicate the size of the
dark matter halo in which the lensing galaxy is embedded) and projected
distance, as a fraction of the measured effective radius.

Alternatively, one may use the stellar mass fundamental plane to
``predict'' the stellar mass surface density at a given projected
radius for a galaxy with a specified velocity dispersion and effective
radius.  Multiplying these predictions by an adjustable constant, and
minimizing residuals between the measured and predicted stellar mass
surface densities allows one to recalibrate the stellar mass
fundamental plane.  If the correct IMF has been used in deriving the
adopted stellar mass fundamental plane, the adjustable parameter,
$\cal F$, will be unity.

In the present work we construct two stellar mass fundamental planes,
one derived from measurements by Auger et al. (2010) for the Sloan
Lens Advanced Camera for Surveys (hereafter referred to as SLACS)
sample at $z \sim 0.2$ and another derived from measurements by
Sonnenfeld et al. (2013) for the Strong Lensing Legacy Survey
(hereafter referred to as SL2S) sample at $z \sim 0.5$.  Both groups
of authors use Salpeter (1955) IMFs to calculate stellar
mass.\footnote{ A low mass cutoff must be applied to the Salpeter
  (1955) IMF to keep the integrated mass from diverging.  Auger et
  al. (2010) and Sonnenfeld et al. (2013) used a low mass cutoff of
  $0.10 M_\odot$ (T. Treu, private communication).}  We use both
samples to constrain the orientation of the fundamental plane, but
then fit them separately to allow for possible evolution with
redshift.

\subsection{Previous investigations}

Past efforts to ascertain stellar (or alternatively dark) mass
contributions to galaxy masses based on micro-lensing have been
carried out in large part by two groups -- one that builds upon the
work of Kochanek (2004), and another that builds upon the work of
Schechter and Wambsganss (2004).  Other efforts include those of
Mediavilla et al. (2009),
Bate et al. (2011) and Oguri et al. (2014), the last of which uses the
stellar mass fraction derived by the present authors (Pooley et al.
2012) as a constraint in what is otherwise a decomposition into two smooth
components, one stellar and one dark.

The present paper produces results that are somewhat less uncertain,
and somewhat more robust, but differs primarily in that our central
goal is a calibration of the conversion of observed light to stellar
mass rather than determining a dark or stellar mass fraction.

\subsection {Quasars as point sources}

The analysis of optical flux ratio anomalies for a set of quadruply
lensed quasars by Schechter and Wambsganss (2004) was inconclusive.
Including the case of SDSS J0924+0219, they found an implausibly low
mass fraction.  Excluding it gave a double peaked likelihood function.
They were able to produce a single peak by making the {\it ad hoc}
assumption that in every case 50\% of the quasar light came from a
point source and 50\% of the light came from a source too extended to
be micro-lensed.

Chandra X-ray Observatory observations of those same quadruply
lensed quasars indicate that finite size effects may indeed have been
responsible for Schechter and Wambsganss' (2004) failure to extract a clean
signal from the optical flux ratios.  Pooley et al. (2007) found 
for a sample of ten lensed systems that the X-ray deviations from
models with smooth potentials, as measured in magnitudes, were a
factor of two larger than the optical deviations.  They argue that the
optical emission comes from a larger region comparable in size to the
Einstein rings of the micro-lensing stars.

While the size of a quasar's optical emission region relative to that
of stellar Einstein rings may be of interest in its own right, for the
present purpose it is yet another free parameter and a major nuisance.
Fortunately, the X-ray emission appears to emanate from a very
compact, more nearly pointlike region.  In the present paper we use
these X-ray observations to determine stellar mass surface densities.

\subsection{Smoothly distributed dark matter}

The micro-lensing magnification distribution expected at the position
of one of our quasar images depends not only on the stellar mass
density but also on the tidal shear due to the galaxy and the
magnification produced by the smooth dark matter density at that
position.  We use an isothermal model for the potential produced by
the combined gravity of the stellar and dark matter components of our
lensing galaxies, consistent with results from the SLACS survey
(Gavazzi et al. 2007; Auger et al. 2010), the SL2S survey (Ruff
  et al. 2011), and a combined analysis of the SLACS and BELLS samples
  (Bolton et al. 2012).

\subsection{Outline}

In \S2 we construct stellar-mass fundamental planes using the SLACS data of
Auger at al (2010) and the SL2S data of Sonnenfeld et al. (2013) for
subsequent use in analyzing quasar micro-lensing.

In \S3 we assemble the observational data needed for our analysis. 
These include positions and effective radii for the
lensing galaxies, and positions and X-ray fluxes for the lensed quasar
images.

In \S4 we present macro-lensing models for the lensed quasars, with
particular emphasis on the total mass surface densities and shears at
the quasar image positions.

In \S5 we describe the suite of ray tracing
simulations used in our likelihood analysis and our scheme for
interpolating across them.

In \S6 we present the details of our likelihood analysis.

In \S7 we carry out that likelihood analysis and derive a re-calibration
factor $\cal F$ for the SL2S stellar-mass fundamental plane.
We then examine and discuss possible sources of systematic errors:
our choice of models for the macro-lensing potential, 
our implicit prior on the unlensed x-ray fluxes of the lensed quasars,
our method of analysis, systematic errors in the measurement of 
effective radii, and the sensitivity of our combined results to 
the results for the individual lensed systems.

In \S8 we discuss our result in the context of other measurements.

In \S9 we summarize our results and describe avenues for further
refinement.

\section{The Stellar Mass Fundamental Plane (and Line)}

\subsection{Basic idea}

The stellar mass fundamental plane (Hyde and Bernardi 2009) differs
from the conventional fundamental plane (Djorgovski and Davis 1987)
in that stellar surface mass density replaces stellar surface
brightness as one of the three ``observables'' measured for elliptical or
early-type galaxies.  In either case the galaxies lie very close to a
two-dimensional planar surface in the three-dimensional space spanned
by the logarithms of the observables. 
For a given stellar mass fundamental plane
one might expect different (parallel) conventional fundamental 
planes for populations of different ages and metallicities.
Observations of the galaxies with several filters would permit
one to combine galaxies of different ages and metallicities into a
single stellar mass fundamental plane.  But this can be
accomplished only by making assumptions about the initial mass function,
since the stars that constitute the bulk of the mass contribute very
little starlight.

Hyde and Bernardi cast their fundamental plane as a ``prediction'' of
the effective radius $r_e$ of an elliptical galaxy,
assumed to follow a de Vaucouleurs (1948) surface brightness profile,
given measurements of 
the average stellar surface mass density $\Sigma_e$ interior
to the effective radius and the stellar
velocity dispersion, $\sigma$,
\begin{equation}
\log r_e = \alpha \log \sigma + \beta \log \Sigma_e + \gamma \quad .
\end{equation}
Observed values of the effective radius (the geometric mean
of the semi-major and semi-minor axes) are compared with the
predictions obtained with trial values of the coefficients, which are
then adjusted to minimize the scatter between the observations and
predictions.

For the present purposes we want a fundamental plane that ``predicts''
stellar surface mass densities as a function of effective radius
and velocity dispersion, 
\begin{equation}
\log \Sigma_e = a \log \sigma + b \log r_e + c \quad .
\end{equation}
We use the predicted surface mass densities to calculate the
effects of micro-lensing at the four quasar image positions, and then
adjust the constant $c$ to maximize the likelihood of the observed
fluxes based on the micro-lensing predictions.  This is the sense
in which we re-calibrate the stellar mass fundamental plane.  We
start with a fundamental plane computed from multi-color observations
and adjust the stellar mass zeropoint to maximize the likelihood of
the observed flux ratio anomalies.

\subsection{Einstein ring radius as a proxy for stellar velocity dispersion}

The use of the stellar mass fundamental plane to estimate stellar mass
surface densities is rendered more challenging by the fact that our
lensing galaxies are crowded by four bright quasar images, making it
either difficult or impossible to measure stellar velocity dispersions
for many of our lensing galaxies.\footnote {Despite this observational
  challenge, PG 1115+080, HE 0435-1223, RX J1131-1231 and RX J0911+0551
  do have measured velocity dispersions} Fortunately lensing galaxies
are characterized by an Einstein ring radius that has a
straightforward relation to the stellar velocity dispersion.  We
follow Kochanek et al. (2000) in constructing a stellar mass
fundamental plane that employs these Einstein ring radii, converted to
equivalent stellar velocity dispersions, and refer to them as
``proxy'' dispersions.

The potentials of
early-type elliptical galaxies are very nearly isothermal (e.g. Auger
et al. 2010), and for an isothermal sphere the isotropic stellar
velocity dispersion can be read directly from the radius of the
Einstein ring, $\theta_{Ein},$ measured in radians, 
\begin{equation}
\sigma_{prox} = c \sqrt{{\theta_{Ein}\over 4 \pi} {D_{LS}\over D_{OS}}} \quad, 
\end{equation}
where $D_{OS}$ and $D_{LS}$ are the angular diameter distances from
the observer to the source and from the lens to the source,
respectively (Narayan and Bartelmann 1997).  The radius of the
Einstein ring, $\theta_{Ein},$ is a direct output of our lens modeling
described in \S4 below.

The stars in the lensing galaxy may not have isotropic orbits.  In
that case, even though the potential might be that of an isothermal
sphere, stellar velocity dispersions will depend upon the aperture
used to measure them.  By contrast the Einstein ring radius suffers
from no such shortcoming.  Finding virtue in  necessity, we use the
Einstein ring radius as a proxy for stellar velocity dispersion in our
construction of a stellar mass fundamental plane.

\subsection{A stellar mass fundamental plane at z $\sim$ 0.2}

We use the data of Auger et al. (2010) for systems of emission line
galaxies lensed by foreground early type galaxies at a mean redshift
of 0.2 to compute a stellar mass fundamental plane.  Auger et al. give
effective radii $r_{e}$ measured in kpc computed at in the rest band
of the $V$ filter, and total stellar masses $M^*$ based on two
different assumed initial mass functions, Salpeter (1955), with a
cutoff of $0.1 M_\odot$, and Chabrier (2003), for 51 systems.\footnote{Following Auger et al. we adopt an $(h, \Omega_m, \Omega_{\Lambda}) = (0.7, 0.3, 0.7)$
  cosmology.}  
Treu et
al. (2010) give Einstein ring radii computed from their singular
isothermal ellipsoid lens models and expressed as proxy velocity
dispersions for the same sample.  As a starting point, we use these
data to calculate a Salpeter stellar mass fundamental plane.

We divide our measured proxy dispersions and effective radii by
typical values, 266 km/s and 6.17 kpc respectively, so as to reduce
(but not completely eliminate) the covariances between the derived 
coefficients, 
\begin{equation}
\log \Sigma_e = a \log \left(\sigma_{prox} \over 266~{\rm km/s} \right)
+ b \log \left(r_e \over 6.17~{\rm kpc} \right) + c \quad .
\end{equation}
The coefficients $a$, $b$ and $c$ are given in Table 1, as is the $r_{ab}$ correlation coefficient.

\subsection{Stellar mass fundamental plane using stellar velocity dispersions}

We are driven to use Einstein ring radius as a proxy for stellar
velocity dispersion because the light from our quasars overwhelms that
of the lensing galaxies.  But as Auger et al. (2010) are working with 
fainter lensed
background galaxies, they {\it have} been able to measure stellar
velocity dispersions, and their data may be used to calculate a proper
stellar mass fundamental plane for the sake of comparison.  The results
are also given in Table 1.

The $a$ and $b$ coefficients for the stellar velocity dispersion
fundamental plane are only marginally consistent with those obtained
for the proxy dispersion fundamental plane, but the range of
dispersions in our sample is relatively small.  As the stellar
velocity dispersions are slightly larger than the Einstein ring
proxies, the zeropoint $c$ is correspondingly smaller.

Also shown in Table 1 is the rms scatter in the logarithm 
of the surface mass density.
It is considerably smaller, by a factor of 1.5, when one
uses the proxy dispersion rather than the stellar dispersion.  The
most likely cause is measurement errors in the stellar dispersions,
which are of order 10\%.  By contrast Einstein ring radii are accurate
to 1\%.

It might also be the case that {\it both} the stellar velocity
dispersion {\it and} the Einstein ring radius are stand-ins for some
third quantity which gives a yet tighter fundamental plane.  If that
third quantity were the size (or mass, or energy per unit mass) of the
halo, one might expect the radius of the Einstein ring to represent it
more faithfully than the stellar velocity dispersion, as it is less
influenced by the stellar mass density at the center of the galaxy.

\subsection{A stellar mass fundamental plane at z $\sim$ 0.5}

The data presented by Sonnenfeld et al. (2013) for lenses found in the
SL2S may also be used to construct a stellar mass fundamental plane.
While only 21 of their systems have complete data, the median redshift
for their sample, 0.494, is much closer to the median for the lens systems
used here.
The sample was subject to different selection effects from
that of Auger et al. (2010) but a common method for stellar mass estimation
was used for both samples.  A different scheme was used for measuring
effective radii but they have also used singular isothermal
ellipsoids to model their lenses and derive proxy velocity
dispersions.

The coefficients derived for the SL2S sample are consistent
with those derived for the SLACS sample but roughly a factor of three
more uncertain.  The scatter between the predicted and observed
surface mass densities is higher for SL2S than it was for SLACS
and, by contrast, no better for the proxy velocity dispersions 
than for the actual measured stellar velocity
dispersions.

\subsection{The orientation of the stellar mass fundamental plane}

So as to better compare the surface mass density at different
redshifts, we wish to adopt a single orientation for the fundamental
plane and fit that to both the SLACS and the SL2S.  We have combined
velocity dispersion and effective radius coefficents for the SLACS and SL2S
samples using the $a$ and $b$ coefficients and uncertainties given in Table 1,
to produce $a$ and $b$ coefficients for the combined sample.  These
are also given in Table 1.

\subsection{Possible differences between z $\sim$ 0.2  and z $\sim$ 0.5}

We obtained new values for the stellar surface mass density coefficients 
for the SLACS and SL2S samples fixing the $a$ and $b$ coefficients
at the common value.  These are too are given in Table~1.

We see that stellar mass surface density at fixed velocity dispersion
and effective radius for the SLACS sample is marginally higher, by
roughly 11\% than for the SL2S.  While the difference might be due to
different selection criteria for the two samples, or to the different
techniques used to extract effective radii, it might also be due to
evolution in the fundamental plane.  We therefore take the common
orientation SL2S stellar mass fundamental plane as our primary
predictor of the surface mass density along the lines of sight to our
quasar images.

We note that Bezancon et al. (2013) see little or no evidence for evolution
in the conventional stellar mass fundamental plane.

\subsection{Systematic errors in effective radii} 

Our stellar mass fundamental plane is only as good as the effective
radius measurements, the velocity dispersion measurements (or their
proxies), and the stellar mass estimates (derived from stellar colors)
used to construct it.  We have framed our effort as an attempt to
calibrate the SLACS and SL2S stellar masses.  But if there were a
systematic error in the SLACS or SL2S effective radii, one might
expect a corresponding systematic error in that calibration.

The situation is complicated by the fact that the effective
radius measurements and the stellar mass surface density
measurements are correlated.  Unless the lines along which
these quantities are correlated lie within the error-free 
stellar mass fundamental plane, the orientation of the
derived plane will deviate systematically from the error-free orientation.

\subsection{A fundamental line for early type galaxies}

One can construct a ``fundamental line'' for either of our samples by
fitting separately for effective radius as function of velocity
dispersion and stellar surface mass density as a function of velocity
dispersion.  Together these two relations give a line in the three
dimensional space of observables.\footnote{Nair, van den Bergh and
  Abraham (2011) describe a related fundamental line in the two
  dimensional space spanned by effective radius and luminosity.
  Insofar as stellar mass and luminosity are strongly correlated,
  their fundamental line is a projection of the present one.}  
Both
the effective radii and the stellar surface mass densities scatter
about their respective mean relations, but that scatter is strongly
correlated, so that the galaxies fan out from the line into a plane.

Keeping that correlated scatter in mind, the relations
\begin{equation}
\log \Sigma_e =  A \log \left(\sigma_{prox} \over 265~{\rm km/s} \right) + C \quad {\rm and} 
\end{equation}
\begin{equation}
\log r_e = D \log \left(\sigma_{prox} \over 265~{\rm km/s} \right) + E \quad ,
\end{equation}
can be used to look for possible systematic errors in the effective
radii measured for the galaxies lensing our quasars, and to
corroborate the result we obtain with the stellar mass fundamental
plane.  The denominator within the parenthesis is the sample
logarithmic mean for the proxy velocity dispersion.

As we intend to use this only as a check on our fundamental 
plane results, we present in Table 2 only results for fits
to the SLACS data.  The uncertainties in the coefficients
derived for the smaller SL2S sample renders it less useful
for our purposes.

\section{Data for Lensed Quasars and Lensing Galaxies}

\subsection{Choice of sample}

Our sample of ten multiply imaged quasars is a subset of the fifteen
systems considered by Blackburne et al. (2011).

The system Q2237+030 was eliminated because the half-light radius of the
lensing galaxy (at $z = 0.04$) is very much bigger than the Einstein ring
radius, and the system appears to be a barred galaxy.  Not
coincidentally it is the system that is best suited to a
multi-epoch analysis, and has been analyzed by Kochanek (2004).

The system WFI J2026-4536 was eliminated because the redshift 
of the lensing galaxy has not been measured.

Two more systems, HE 1113-0641 and H1413+117, were eliminated because 
the lensing galaxy is barely detected and does not yield 
a reliable effective radius.  

The system SDSS J1330+1810 was eliminated because X-ray fluxes have
not been measured.  

Of the original fifteen systems studied by Blackburne et al. (2011),
ten systems remain in the sample.

\subsection{X-ray fluxes}

X-ray fluxes and errors were taken,
for all but two cases,
from Pooley et al. (2007),
who chose from among multiple epochs based on 
signal to noise.  For the two subsequently observed systems we used 
the X-ray fluxes and errors given by Blackburne et al. (2011). 
These selection criteria ought not to bias the results derived
here. The dates and identification numbers
for Chandra observations are given in Table 3.
Fluxes for the individual images are given in Table 4.

\subsection{Effective radii}

The largest homogeneous source of effective radii for our quasar lenses
is a series of papers that use Magain, Courbin and Sohy's (1998;
henceforth MCS) algorithm 
(Claeskens et al. 2006;
Eigenbrod et al. 2006; 
Vuissoz et al. 2008;
Chantry et al. 2010;  
Courbin et al. 2011; 
Sluse et al. 2012a).
The second largest is the series by Kochanek and
collaborators (Kochanek et al. 2000, 2006; Morgan et al. 2006) which uses a program called {\tt imfitfits} (Lehar et al. 2000). 

Measurements using the two methods are given in Table 5.
There are five systems in common.  The MCS values are consistently larger than the
{\tt imfitfits} values by a factor of 1.62 (computed by averaging logarithms
of ratios).  
It would appear that {\tt imfitfits} systematically underestimates the
effective radii, or that the MCS method systematically overestimates the
effective radii, or perhaps both.
We have found no persuasive reason to prefer one over the other.
We have therefore taken a ``split-the-difference" approach,
adopting the geometric mean when we have measurements with both methods. 
Where we have only an {\tt imfitfits} measurement we multiply by $\sqrt{1.62}$
and where we have only an MCS measurement we divide by $\sqrt{1.62}$.

For the one case in which we have independent measurement, that of
SDSS J0924+0219, we adopt the effective radius measurement of Keeton
et al. (2006) who describe a method much like that used in the SLACS
survey.  For HE 0230-2130 we present newly determined effective radii
(Schechter and Levinson, in preparation) and ellipticities and
position angles measured with {\tt GALFIT} (Peng et al. 2002) from
images obtained at the Magellan Clay telescope.

The SLACS effective radii were obtained from fits of an elliptical de
Vaucouleurs profile to a masked image of the lensing galaxy (Bolton et
al. 2008).  The SL2S effective radii were computed using {\tt GALFIT}.
Since we are interested in calibrating stellar mass fundamental planes
obtained from these surveys, it behooves us to check whether our
adopted effective radii are consistent with these.

In Figure 1 we plot the adopted effective radii against proxy velocity
dispersions computed from Einstein ring radii (see \S2.2 and \S4.2)
for the present sample (filled circles) and for the SLACS sample (open
circles).  There is considerable overlap but considerable scatter.
The overlap might be somewhat greater if we adopted yet larger
effective radii.  We note, however, that overlap with the SLACS and
SL2S relations is no guarantee that our adopted effective radii are
consistent.  It is possible that our lensing galaxies occupy a
different part of the stellar mass fundamental plane than the SLACS
and SL2S samples, with a corresponding shift in effective radii.  In
\S7.4 we discuss the possible effects of a systematic error in our
adopted effective radii.

\subsection{Positions}

Positions for the quasar images and lensing galaxies given in Table 4
are taken from Blackburne et al. (2011), with the exception of those
for B1422+231, which are taken from the CASTLES gravitational lens data base.\footnote{http://www.cfa.harvard.edu/castles}

\subsection{Lens galaxy ellipticities}

Ellipticities and position angles for the lensing galaxies are given
in Table 4.  All were taken from the compilation by Sluse et
al. (2012a) in their table B.2, with the exception of those for HE 0230-2130,
which is again from Levinson and Schechter (in preparation).

\section {Macro-Models for Lensing Potentials}

Our calibration of the stellar mass fundamental plane relies on the
use of micro-lensing simulations of the macro-images.  The
simulations, described in \S5 below, are parameterized by a
convergence, $\kappa$, a shear, $\gamma$ and a stellar contribution to
the convergence, $\kappa^*$.  The appropriate
convergence and shear are determined from macro-models for the lensing
potentials.

\subsection{Singular isothermal ellipsoid + external shear}

In our previous work (Blackburne et al. 2011; Pooley et al. 2012) we
adopted models for the gravitational potential comprised of one or
more singular isothermal spheres representing lensing galaxies at
their observed positions and an additional external tidal shear.  We
refer to these as ``SIS+X'' models.  The choice was motivated
foremost by the observation that for several of our systems there was
an obvious source for the deduced external tide.

For several other systems a model in which the quadrupole component
of the potential originated in the lens, the singular isothermal
ellipsoid (henceforth SIE), provided a comparably good (but not better) fit.  In
the interest of simplicity Blackburne et al. (2011) adopted SIS+X
models for all of the systems considered.  Additional justification
for this choice came from the work of Kochanek (2006), who found that
external shear dominates in lensing systems.

This introduces a bias into our derived stellar mass fractions.  The
SIE models typically have smaller magnifications than
the SIS+X models.  The higher the magnification of a point source, the
larger the number of micro-minima introduced by the micro-lensing stars
(Granot et al. 2003).  The expected fluctuations are therefore
different in the two cases.

In the present study we model the potentials with both a singular
isothermal ellipsoidal mass distribution {\it and} an external shear
-- SIE+X models. 
The ellipticity and orientation of the SIE -- comprised mostly of dark
matter -- are not treated as free parameters, but are
taken from the observed shape and orientation of the stellar
component.  Our SIE+X models therefore have the same number of free
parameters as the Blackburne et al. (2011) SIS+X models.

The assumption that the dark halo has the same shape as the luminous
stellar component is a strong one.  While this is roughly consistent
with previous findings (Koopmans et al. 2007), we introduced it to
eliminate potential bias rather than to improve residuals from our
models.  But in seven out of ten cases, the residuals from the
observed positions of the quasar images were substantially better for
the SIE+X models than for the SIS+X models.  For the three cases in
which the SIE+X models gave worse fits, we use the Blackburne et al.
SIS+X models.  The present models are quite similar to the Blackburne
et al. (2011) models, but with the macro-images slightly less strongly
magnified.  The systematic effect of this difference on our derived
surface mass densities is discussed in \S 7.7.

Keeton's {\tt lensmodel} (2001) program was used to obtain the model
parameters that give a best fit to the observed quasar image
positions.  The fits were not constrained by the quasar fluxes since
we expect deviations from the macro-models due to micro-lensing;
these deviations are the signal we use to determine stellar masses.

The model parameters from our fits are given in Table 6.  The results
are very similar to those in Blackburne et al. (2011), but the
systems for which we
used the SIE+X model have slightly smaller magnifications.
The derived convergences and shears, computed at the predicted (model)
positions for the quasar images rather than those observed, are given
in Table 4.  The designations LM, LS, HM and HS indicate whether the
images are minima (M) or saddle points (S) of the light travel time,
and whether they are more highly magnified (H) or less highly magnified
(L).\footnote{
We originally adopted this notation to indicate minima and saddlepoints
that were higher and lower on the light travel time surface.  But in the
case of HE 0230-2130 the higher saddle point is less highly magnified.}

The model for HE 0230-2130 as two isothermals differs from that
  of Pooley et al. (2007) who modeled the density of the second galaxy
  as a power law, $\rho = \rho_0(r/r_0)^{-\gamma'}$ with an exponent
  $\gamma' = 1.65$.  The latter gives magnifications that are larger
  than the former, by factors of 1.25 for images $a, b,$ and $c$
  and 1.7 for image $d$.  In \S7.4 we show that use of the Pooley et
  al. (2007) model produces negligible change in our principal result.

  But our double isothermal model produces an unobserved fifth image, a saddle
  point (Shin and Evans 2008), 
  which is found where one would expect an image if the two
  lensing galaxies were to merge, adjusting the quadrupole to keep
  $a,~b$ and $c$ at approximately the same positions.  The image
  should be as bright as image $c$ and is clearly not present. While
  the Pooley et al. (2007) model produces no fifth image, it is {\it ad
  hoc}.  Both models attribute a much smaller velocity dispersion to
  the second galaxy than would be warranted by its absolute magnitude.
  A more satisfactory model awaits measurement of mid-IR or forbidden
  line flux ratios, as discussed in \S7.8.

\subsection{Proxy velocity dispersions}

Our models give Einstein ring radii measured in arcseconds.  Using the
lens and source redshifts, we compute the proxy velocity dispersions
described in \S2.2 that we need for our our stellar mass fundamental
plane.  These are given in Table 7.

\section{Micro-lensing simulations}

Our strategy is to use the stellar mass fundamental plane to predict
deviations in the observed quasar X-ray fluxes from the predictions of
the smooth models for the gravitational potential.  As there is no
closed form prediction for such fluctuations we must simulate the
observed quasar images.  We use simulations carried out by Blackburne
et al. (2011) using the ray-shooting technique described by Wambsganss
(1990).  

The Blackburne et al. (2011) simulations used the convergences and
shears appropriate to the image positions in their SIS+X models, and a
range of stellar contributions at each position.  They populate a
three dimensional model space spanned by $\kappa, \gamma,$ and $\kappa^*.$

Our SIE+X models have somewhat different convergences and shears from
the SIS+X models.  But since the magnification histograms derived from
these simulations vary smoothly as one varies the parameters, we have
chosen to interpolate between the existing magnification histograms
rather than run new simulations.

Our interpolation scheme takes advantage of the well known mass sheet
degeneracy.  For any triplet of convergence, $\kappa$, shear,
$\gamma$, and stellar contribution to the convergence, $\kappa^*$,
there is a one dimensional locus -- a line of triplets 
$(\kappa', \gamma', {\kappa^*}')$ -- in
the three dimensional model space for which the magnification
histogram has the identical shape but is shifted by a multiplicative
factor (Paczy\'nski 1986a; Kochanek 2004; Vernardos et al. 2014).  
This family is compactly parameterized by the 
the smooth (and in our case dark) contribution to
the convergence, $\kappa_s \equiv \kappa - \kappa^*$,
\begin{eqnarray}
{\kappa^*}' & = &\kappa^* \left({1 - \kappa'_s \over 1 - \kappa_s}\right) \quad ,\\
\gamma' & = &\gamma \left({1 - \kappa'_s \over 1 - \kappa_s}\right) \quad, \\
(1 - \kappa') & = &(1 - \kappa) \left({1 - \kappa'_s \over 1 - \kappa_s}\right) \quad .
\end{eqnarray}
Projected onto the two dimensional convergence-shear,
$(\kappa',\gamma')$, plane the family lies along the line connecting the
point $(\kappa, \gamma)$ to the point with unit convergence and zero
shear, $(1,0)$.

Our lens models, and those of Blackburne et al. (2011) give values for
the convergences and shears (Table 4) that cluster around the line $\kappa
= \gamma$,  which is exactly the case for
an unperturbed isothermal sphere.  Our interpolation
scheme is then to project all of the simulations onto the line
$\kappa' = \gamma'$, with corresponding values of ${\kappa^*}'$.  Our
macro-model and the stellar mass fundamental plane gives us model
values for the convergence, shear and stellar contribution to the
shear $\kappa_{mod}, \gamma_{mod}$ and $\kappa^*_{mod}$.  We also project
these onto the $\kappa' = \gamma'$ line.   We linearly interpolate
between the bracketing simulations, and then linearly interpolate these
between bracketing values of ${\kappa^*}'.$ This gives us a
magnification histogram appropriate to $\kappa_{mod}, \gamma_{mod}$
and $\kappa^*_{mod}$.
 
In calibrating our stellar mass fundamental plane we explore a wide
range of multiplicative factors.  There are some cases, at the
extremes of this range, for which $\kappa^*_{mod}$ lies outside the range
of the Blackburne et al. (2011) simulations.  In such cases we have used the
closest simulation.  All of these were at very low likelihood, and do
not significantly affect our conclusions.

\section{Likelihood analysis}

Our goal is to obtain a ``best'' estimate of the factor $\cal{F}$ by which
the surface mass density predictions from our stellar-mass fundamental
plane must be multiplied to reproduce the micro-lensed X-ray fluxes observed
for our quasar images.   

\subsection{Stellar surface mass density predictions}

Given the proxy velocity dispersion $\sigma_{prox}$ computed from the
Einstein ring radius of the lensing galaxy and the photometric
effective radius $r_e$ of that galaxy, one can compute the predicted
surface mass density at the position of the $i^{th}$ quasar image.

Let $u_i$ and $v_i$ be the position of the $i^{th}$ image relative
to the center of the lensing galaxy, with the $u$ and $v$ axes aligned
with the major and minor axes of the lens, respectively.  For a galaxy
with ellipticity $e$, contours
of constant surface brightness lie along a locus of constant $w_i$,
with 
\begin{equation}
w_i^2 = (1 - e) u_i^2  + v_i^2/(1-e) \quad . 
\end{equation}
The surface mass density is computed using de Vaucouleurs' (1948) law,
\begin{equation} 
\Sigma (u_i, v_i) = {\cal F}{\Sigma_e \over 3.607}
\exp \left[ -\left(w_i\over r_e \right)^{1 \over 4} + 1 \right] \quad ,
\end{equation} 
where $\Sigma_e$ is the average surface mass density {\it interior} to the
effective radius computed from the stellar mass fundamental plane, which
divided by 3.607 gives the surface mass density {\it at} the effective
radius.
In what follows the dimensionless factor ${\cal F}$ is varied to
maximize the likelihood of the observed quasar fluxes.

The stellar contribution to the dimensionless convergence, $\kappa^*,$
is then given by 
\begin{equation} 
\kappa^* = {4 \pi G \over c^2} {D_{OL} D_{LS} \over D_{OS}} 
\Sigma(u_i,v_i) \quad, 
\end{equation}
where $D_{OL}, D_{LS}$ and $D_{OS}$ are angular diameter distances.

While several of our systems have nearby perturbing galaxies,
for all but one of them the stellar surface density may be
taken to come entirely from the central galaxy.  The exception is HE 0230-2130,
for which a second lensing galaxy lies just beyond image $d$.
For this system we take the sum of the stellar contributions
from the two galaxies.

\subsection{Probability of micro-lensed flux}

For each quasar there is an unmagnified flux that may be expressed
as a magnitude, $m_s$.  The macro-model gives  macro-magnifications
$\mu_i$ for each image $i$, computed from the convergences $\kappa_i$
and shears $\gamma_i$ given in Table 4,
\begin{equation}
\mu_i = {1 \over [(1-\kappa_i)^2 - \gamma_i^2]} \quad .
\end{equation}
For each image, the flux expected from the macro-model, expressed
as a magnitude is 
\begin{equation}
\bar m_i \equiv m_s -2.5\log \mu_i \quad . 
\end{equation} 
Taking $m_i$ to be the micro-lensed flux, expressed as a magnitude, for
image $i$, a micro-lensing magnification histogram gives us the relative probability
of observing an offset from the flux expected from the macro-model alone, 
${\cal P}(m_i - \bar
m_i | \kappa_i, \gamma_i, \kappa^*_i)$ where $\bar m_i$ depends upon
the unmagnified flux and the macro-model for the lensing potential and
where the stellar contribution to the convergence, $\kappa^*_i$,
depends upon the adopted stellar mass fundamental plane, the observed
proxy velocity dispersion, effective radius and ellipticity, and the scaling
factor ${\cal F}$.

\subsection{Likelihood of unmagnified flux}

We take the likelihood $\cal L$ of the unobservable unmagnified flux, $m_s$,
to be  the probability of the micro-lensed flux for the $i^{th}$ image,
$m_i$, giving us 
\begin{equation}
{\cal L}(m_s | m_i, \kappa_i, \gamma_i, \kappa^*_i) =
{\cal P} (m_i - m_s + 2.5\log \mu_i | \kappa_i, \gamma_i, \kappa^*_i) \quad .
\end{equation}
Note that we have implicitly assumed a uniform prior on $m_s$ the unmagnified
flux expressed as a magnitude.

For any given observation, the observed flux
for the $i^{th}$ image, $m_{o,i}$, is comprised of the
micro-lensed flux $m_i$ plus an observational error, $\Delta_i$. 
Taking the observational 
uncertainty in the observed
X-ray flux, expressed as a magnitude, to be a Gaussian
of width $\sigma_i$, and integrating over
the difference $\Delta_i$ between the observed and micro-lensed fluxes
we find 
\begin{equation}
{\cal L}(m_s | \kappa_i, \gamma_i, \kappa^*_i) =
\int \exp\left[-{\Delta^2_i \over 2 \sigma_i^2}\right]
{\cal L}(m_s | m_{o,i} - \Delta_i, \kappa_i, \gamma_i, \kappa^*_i) d\Delta_i \quad .
\end{equation}
where we have ignored the small asymmetries in the flux errors.
The integral has the effect of smoothing the magnification
histogram.  

Taking the product of the likelihoods for the four images of one
of our systems we have 
\begin{equation}
{\cal L}(m_s | {\cal F}) = 
\prod_{i=1}^{i=4} {\cal L}[m_s | \kappa_i, \gamma_i, \kappa^*_i({\cal F})] 
\quad,
\end{equation}
where we now show explicitly the dependence of the four stellar contributions
to the convergence, $\kappa^*_i$, on the factor ${\cal F}$ that calibrates
the stellar mass fundamental plane.

\subsection{Likelihood of the stellar mass calibration factor ${\cal F}$}

We take the likelihood of the stellar mass calibration factor ${\cal F}$
to be the integral of the likelihood of the unobservable unmagnified
X-ray flux, 
\begin{equation}
{\cal L}_j({\cal F})= 
\int {\cal L}_j(m_{s,j} | {\cal F}) dm_{s,j} \quad ,
\end{equation}
where we have added the subscript $j$ to emphasize that this
is for the $j^{th}$ lens in our sample.  Were we to place a prior
on the $m_{s,j}$, it would appear inside the integral on the right.

Finally we take the product of the likelihoods for our ten systems to give
\begin{equation}
{\cal L}({\cal F}) = 
\prod_{j=1}^{j=10} {\cal L}({\cal F}_j) \quad,
\end{equation}
to give the likelihood of the calibration factor for the stellar mass 
fundamental plane, $\cal F$, for our complete sample.

\section{The calibration factor $\cal F$ for the stellar mass fundamental
plane}

\subsection{Citable result}

We have constructed the likelihood function $\cal L(F)$ for the
calibration factor that multiplies our adopted stellar mass
fundamental plane based on a Salpeter initial mass function
using the SL2S stellar masses computed by Sonnenfeld et al. (2013).

In Figure 2 we show the relative likelihood for different values of
that factor, increasing in steps of $2^{1/4}$ ($ \approx 1.189$).
This more than doubles the sampling of our initial grid of
simulations, which were carried out in steps of $10^{1/6}$ ($ \approx
1.468$) in stellar surface density.  The median likelihood value is
${\cal F} = 1.23$ with a 68\% confidence interval given by $0.77 <
{\cal F} < 2.10$.

\subsection{Errors in X-ray fluxes}

We have investigated the consequences of having underestimated our
errors by doubling the errors in Table 4 after first expressing them as
errors in the logarithm of the flux.  We find a shift
in the median  of $\cal F$ to a value 24\% higher, which is still comfortably
inside the original 68\% confidence interval.

\subsection{Redshift evolution of stellar mass fundamental plane}

Bezancon et al. (2013) have argued that there is relatively little
evolution in the stellar mass fundamental plane with redshift.  But as
noted in \S2.5, the SLACS and SL2S samples, with median redshifts of
0.205 and 0.494, respectively, give slightly different surface
mass densities if one assumes the same orientation for the stellar
mass fundamental plane at both redshifts.  This might indicate a
systematic difference between the two samples in $\Sigma_e$, $r_e$ or
$\sigma_{prox}$, or it might alternatively be taken to indicate an evolution
in the calibrating factor $\cal F$.  We report the result obtained
using the SL2S sample because it lies much closer in redshift to the
lensing galaxies in the present sample.

\subsection{Systematic errors in lensing galaxies' effective radii}

 As noted in section \S3.3 above, there is a systematic difference
  between effective radii measured using {\tt imfitfits} and the MCS method,
  with the latter larger than the former by a factor of 1.62.  We have
  used the geometric means of the two results when both were
  available, and multiplied the {\tt imfitfits} results by $\sqrt{1.62}$ and
  divided the MCS results by $\sqrt{1.62}$ when only one was
  available.  We have tested the effect of a systematic error in our
  adopted effective radii by increasing them all by a factor of
  $\sqrt{1.62}$.  These give a calibrating factor ${\cal F}$ 23\%
  smaller than our fiducial value.

We take this smaller factor to result from the fact that
while the stellar mass fundamental plane gives smaller surface
densities at the MCS effective radii, this is more than compensated by the
fact that the quasar images are then less far out on the rapidly
declining de Vaucouleurs profile.

The large differences in the measured effective radii for the lensed
quasars may result from the difficulty of measuring the surface
brightness profile of the lensing galaxy in the presence of four
bright quasar images.  

We can completely circumvent effective radius measurements
by using the ``fundamental line'' relations obtained in \S2.9.
The lens model gives us the proxy stellar velocity dispersion $\sigma_{prox}$.
We then obtain effective radii from the $r_e(\sigma)$ relation and
the effective surface mass density from the $\Sigma_e(\sigma).$ 

The scatter from the fundamental line is somewhat larger than from the
fundamental plane, and there is some danger that our sample deviates
from it systematically.  It is therefore reassuring to see that the
median likelihood for our calibration factor $\cal F$ obtained using
the fundamental line is  only 7\% larger than that obtained using
the fundamental plane.  Moreover the likelihood distribution is if
anything, narrower.

\subsection{Systematic errors in SLACS or SL2S effective radii}

A systematic error in either the SL2S or SLACS radii manifests
itself as a shift in both the effective radii, $r_e$, by some factor
$f$, and in the observed stellar surface mass densities, $\Sigma_e$.
Since the effective stellar surface mass densities
are obtained from total magnitudes divided by $r_e^2$.  
Such an error would drive the constant $c$ in the stellar mass fundamental plane down by
a factor $1/f^2$.  But since $\Sigma_e \sim r_e^{-1.453}$, the
net factor by which the  predicted surface densities are smaller
is  $1/f^{0.547}$.

\subsection{Effects of mass sheet degeneracy}

The well known mass sheet degeneracy permits the addition of a uniform
surface density mass sheet to a lens model that, with corresponding
adjustment of the model parameters, produces the same image positions
and relative magnifications.  To the extent that lensing galaxies lie
in clusters of galaxies, the cluster dark matter along the line of
sight to the lens acts as such a mass sheet.

In one of our lenses, PG 1115+080, we have taken this into account
explicitly by modeling the associated group of galaxies as an
isothermal sphere, contributing an additional convergence at the
position of the lens galaxy of $\Delta \kappa = 0.105$.  The convergence for an
isothermal sphere is equal to the shear, so to gauge the possible
effect of the mass sheet degeneracy, we might add a convergence,
$\Delta \kappa$, comparable to the external shear that we measure.

Such an additional smooth contribution to the convergence does not
change the magnification histogram.  It does, however, increase the
observed size of the Einstein ring, by a factor $(1 - \kappa)/(1 -
\kappa - \Delta \kappa)$.  Our proxy velocity dispersion is taken to
be a property of the lensing galaxy as opposed to the galaxy plus mass
sheet system.  In the presence of a mass sheet we then overestimate
this as well.

Our convergence values cluster around $\kappa \approx 1/2$, as
expected for an isothermal sphere.  The observed Einstein ring radii
therefore overestimate the velocity dispersion in the galaxy by a
factor $\approx (1 - 2 \Delta \kappa)$.  We see from Table 1 that for
the same measured effective radius we will get a smaller predicted
surface mass density, by a factor $\approx (1 - 2 \Delta
\kappa)^{1.748}$.  These would lead to larger values of the
calibration factor $\cal F$.

While it might seem to be more appropriate to use the smaller Einstein
ring radii, we note that both the SLACS and SL2S radii were derived
assuming no mass sheet.  Since we seek to apply a calibrating factor
to a fundamental plane derived from these data, it would seem best to use
an Einstein ring radius likewise uncorrected for a possible mass sheet.  

There is, however, some reason to think (Holder and Schechter 2003)
that quadruply imaged quasars experience stronger shear than the SLACS
and SL2S lensing galaxies.  Koopmans et al. (2006) place a limit on the
external shear for a subset of the SLACS lenses of 0.035.  By contrast,
we see in Table 6 that the typical shear for our lensed quasars is
0.1.  On the hypothesis that the external shear is the result of a
larger isothermal cluster, the additional convergence would be larger
for the present sample than for the SLACS sample.

The effect would not be large except for the case of RX J0911+0551,
for which the lensing galaxy is clearly part of a cluster of galaxies.
However the shear, with $\gamma = 0.294$, seems not to be directed to the
center of the cluster (Kneib, Cohen and Hjorth 2000).

\subsection{Systematic errors in lens model and QSO magnifications}

In modeling the expected fluctuations one must specify a convergence,
$\kappa$, and a shear, $\gamma$ at the positions of the quasar images.
These depend upon the particular model for the gravitational lens
potential.  Our adopted SIE+X model is a singular isothermal ellipsoid
with ellipticity and position angle taken from optical observations,
with an external source of tidal shear providing as much if not more
quadrupole than the SIE.  

A commonly adopted alternative (and the one explicitly adopted by the
SLACS and SL2S groups) is to attribute all of the quadrupole to an
SIE.  While there are several systems for which the SIE is manifestly
inferior (there are obvious sources of tides) we have constructed SIE
models for our systems and use them to gauge how large a systematic error
we might be making in adopting our SIE+X models.  For the SIE models
we find a median likelihood value for the factor by which the Salpeter
stellar mass must be multiplied which is lower by roughly 17\% than for the
SIE+X models.

We have also produced models with a steeper than isothermal mass density
profile, $\rho(r) = \rho_0(r/r_0)^{-\gamma'}$, with exponent $\gamma' =
2.1$.  The resulting calibration factor is higher by roughly 32\%.

\subsection{Unmodeled high order contributions to the lensing potential
and milli-lensing}

The macro-models for our lenses are simple as a matter of necessity:
they are constrained by the positions of the quasar images but {\it
  not} the fluxes because the latter may be subject to the very
micro-lensing we wish to explore -- we would be running the risk of
modeling out the micro-lensing.  With so few constraints we cannot
allow for more than a quadrupole moment in our lensing galaxies (and
at that we insist that they have the shape and orientation of the
stellar component).  We include the effects of visible companions and
satellites but not unseen dwarf companions.

It is generally thought that the radio, mid-IR, and forbidden line
emitting regions of quasars are large compared with the Einstein rings
of stars and therefore not subject to micro-lensing (Mao and Schneider
1998; Chiba et al 2005; Sugai et al. 2007; but see Sluse et al. 2013
for arguments to the contrary regarding the mid-IR).  Deviations of radio,
mid-IR and forbidden line flux ratios from our models may therefore be
used to gauge the contribution of unmodeled higher order components
of the lensing potentials (henceforth referred to
as ``milli-lensing'') to our X-ray flux ratio anomalies.

In Table 8 we have collected radio 8.4 GHz (Patnaik 1999), mid-IR
$11.7 \mu$ (Chiba et al. 2005; MacLeod et al. 2011] and [O III]
forbidden line (Sugai et al. 2007) flux ratios for four of the lenses
in our sample.  The remaining six lenses are as yet beyond the reach
of these techniques.  Likewise there is no [O III] data for the
faintest image of RX J1131-1231.  We did {\it not} include Fadely and
Keeton's (2011) $L'$ filter observations of HE 0435-1223.  As they
note, the degree to which the $3.8\mu$ emitting region is or is not
subject to micro-lensing depends upon the relative contributions of
the accretion disk (which is presumably micro-lensed) and dusty torus
(which presumably is not) to the observed flux at that wavelength.
Blackburne et al. (2011) found that the flux ratio $b/a$ steadily
increases toward unity going from the optical to the $K_s$ filter at
$2 \mu$.  Fadely and Keeton find that $b/a$ decreases away from unity
going from $2 \mu$ to $3.8 \mu$.  The lens redshift is $z = 1.689$ and
the emitted wavelengths are therefore correspondingly shorter.

As in Table 4, the fluxes in Table 7 are given relative to the
less-magnified minimum, which is in most cases less likely to be
affected by micro-lensing.  We give the ratios (expressed as magnitude
differences) of the observed to the macro-model fluxes for both the
X-ray and alternative data.  The former is presumably affected by both
micro-lensing and milli-lensing; the latter only by milli-lensing.

We note first that the radio, mid-IR and forbidden line fluxes are
very much closer to the macro-model predictions than the X-ray fluxes.
The typical milli-lensing deviation is $\sim 0.25$ mag.  The X-ray
deviations are several times larger, with no obvious correlation in
sign.  These X-ray deviations are the signal that permits our
calibration of the stellar mass fundamental plane.  The milli-lensing
contribution is evidently small.  

The first order effect of milli-lensing would be a shift of the
predicted micro-lensing distribution to brighter or fainter fluxes.
Second order effects would change the shape of that distribution.  The
first order effect of milli-lensing would therefore seem to be the
same as that of measurement errors in the X-ray fluxes.  

To allow for the failure to model milli-lensing we have analyzed our
data but increased the effective uncertainties in our X-ray fluxes by
adding an error of 0.25 mag in quadrature to the purely observational
uncertainties.  The result is to increase $\cal F$ by roughly 5\%.

But rather than just allow for unknown random uncertainties, we can,
for the four systems for which we have radio, mid-IR and forbidden
line data, correct the macro-lens model fluxes by the observed
deviations from those models.  The corrected model constrains the
quantity $1/[(1 - \kappa)^2 - \gamma^2]$ at each image position but
does not constrain separately the convergence $\kappa$ and shear
$\gamma$.  We therefore use the uncorrected convergence and shear to
generate the micro-lensing distribution (as in our uncorrected
calculation) but shift it by the amount indicated by the appropriate
radio, mid-IR or forbidden line observations, for those systems for
which they are available.  For those without such observations
we increase the uncertainties by 0.25 mag as above.
The result is to increase $\cal F$ by roughly 6\%, and interestingly,
to narrow the 68\% confidence interval by roughly 25\%

There is one case, that of HE 0230-2130, where our model has a glaring
shortcoming, as described in \S4.1.  If we adopt as an alternative
the model of Pooley et al. (2007) the calibration factor $\cal F$ is smaller
by 1\%.

\subsection{Variation in the IMF with velocity dispersion}

There is both spectroscopic and dynamical evidence that 
the IMF varies systematically with stellar velocity dispersion.

Spectroscopic evidence for an increasingly bottom-heavy IMF with
increasing velocity dispersion has been presented by van Dokkum and
Conroy (2012), La Barbera et al. (2013) and Spiniello et al. (2014).
These can be used to compute stellar masses, but only if one makes 
the assumption about the low mass cutoff in the IMF (Conroy and van
Dokkum 2012).  Spiniello et al. find an IMF slope consistent with a
Chabrier IMF at a dispersion of 145 km/s, rising to a Salpeter IMF
slope at a dispersion of 240 km/s.  

Dynamical evidence for an increase in the ratio of stellar mass to
that computed from a Salpeter IMF (again with the assumption of a
stellar mass cutoff) has been presented by Treu et al. (2010),
Cappellari et al. (2012) and Conroy et al. (2013).   Smith (2014)
has argued against the interpretation of the spectroscopic
data as the effect of velocity dispersion but accepts the dynamical
argument.

The mean proxy velocity dispersion for our sample is 243 km/s, with
dispersions for the primary effective lensing galaxy ranging from 178
km/s to 349 km/s.  One might adjust the stellar mass fundamental plane
to take variations in the IMF into account, but our range in 
dispersion is too small, our tool too blunt, and our sample (for the
present) too small to confirm such a trend.

\subsection{Variation in the IMF with radius}

There is both direct and indirect evidence that the IMF might vary
with radius.  Mart\'in-Navarro et al. (2014) find that for massive
early type galaxies the IMF is more bottom-heavy toward the center
than further out.  This is consistent with the observation that
stellar abundances within a galaxy vary with escape velocity (Franx
and Illingworth 1990) combined with the dependence of the IMF on
stellar velocity dispersion discussed in the previous subsection.

To first approximation one might take both the stellar mass and the
stellar light to be given by de Vaucouleurs profiles, with the mass
having a smaller effective radius $r_{e,M}$ than that of the light,
$r_{e,L}$.  Using the stellar mass fundamental plane the inferred
average stellar surface mass density interior to $r_{e,L}$ will have
been underestimated, since more than half of the stellar mass will be
interior to this radius.  But he local surface mass density at
$r_{e,L}$ will be a smaller fraction of this average interior density,
since at the effective radius the de Vaucouleurs profile is rapidly decreasing.

Our Einstein ring radii are $\sim 1.5$ times larger than the effective
radii.  If $r_{e,M} < r_{e,L}$, the predicted local surface mass
density will have decreased by a larger factor from $r_{e,L}$ than we
have computed using equation (11), leading to a larger calibration
factor $\cal F$.

\subsection{Prior on unmagnified X-ray fluxes}

Implicit in our likelihood approach is the assumption that all
logarithmic values for the unobservable unmagnified X-ray flux of our
sources ($m_s$ when expressed in magnitudes) are equally likely.  In a
Euclidean universe one would expect the numbers to increase toward
fainter fluxes as $10^{0.6m_s}$.  For optically selected quasars the
numbers first increase more rapidly than $10^{0.6m_s}$, reflecting the
increasing number density of quasars with increasing redshift, and
then more slowly, reflecting a flattening or decline in the quasar
number density at redshifts $z > 3$ and the dominance
of relativistic effects.

Lehmer et al. (2012), in their Figure 9, show numbers of
quasars with $1 < z < 3$ increasing at roughly $10^{0.3m_s}$ in the 0.5
- 2.0 keV band for fluxes $6 \times 10^{-16} < S < 4 \times 10^{-15}$,
which we take to be the relevant range of unmagnified fluxes for our
systems.

As a crude gauge of the effects of selection we have investigated the
effect of a modest prior toward fainter fluxes, $10^{0.3m_s}$.  It
produces an 11\% shift toward a lower calibration factor $\cal F$.

\subsection{Effects of time delay}

Our likelihood analysis involves the implicit assumption that the
unobservable unmagnified X-ray flux is the same for all four images.
The photons that we observe were emitted at different times, spanning
several weeks for most of our systems, but with differences as large
as $\approx$150 days for the case of RX J0911+0551 (Hjorth et al.
2002).  The source may have varied during this time.

The data on X-ray quasar variablility is limited.  Quoting from the
abstract of  Gibson and Brandt (2012)  

{\parindent=20pt
{\narrower
``Assuming that the distribution of fractional deviations is Gaussian,
its standard deviation is $\approx 16$\% on $\gtrsim 1$ week timescales, ...
... extreme variations ($>$ 100\%) are quite rare, while variation at the
25\% level occurs in less that 25\% of observations.''
\par
}
}

Some of the best data on X-ray quasar variability comes from lensed
systems.  Zimmer, Schmidt and Wambsganss (2011) have studied Q2237+0305,
which {\it does} seem to have shown substantial, factor of three
variation on a timescale of $\approx$60 days.

The multiple epoch data for MG J0414+0534, PG 1115+080, RX J1131-1231
and B1422+231 in Table 1 of Pooley et al. (2012) shows only modest
evidence for coherent variation in the X-ray flux quartets, with
substantially larger variations due to micro-lensing.

For the most part the amplitude of the variability is small both with
respect to our observational errors and second with respect to the
width of the micro-lensing histograms, and as such seems unlikely to
influence our results.

\subsection{Differential absorption by intervening matter}

We have investigated the effects of X-ray absorption using the {\it XSPEC}
spectral fitting package (Arnaud et al. 1996).  For a $z=2$ quasar
with an X-ray power law index of 1.7 and a lens at $z=0.5$, a factor
of two reduction in the observed 0.5 - 8 keV band flux requires a
column density of $10^{23} {\rm cm}^{-2}$ (assuming solar abundances) in
the lensing galaxy.
Using the standard dust-to-gas ratio ($N_H/A_V = 1.79 \times
10^{21} {\rm cm}^{-2}{\rm mag}^{-1}$) this corresponds to 55
magnitudes of attenuation in the observed $V$ filter (Predehl \&
Schmitt 1995).  The absence of strong differential reddening at
optical wavelengths argues against absorption as a contributor
to the X-ray flux ratio anomalies.

\subsection{Consistency check on likelihood analysis}

As a check on the consistency of our approach we have generated
synthetic X-ray observations using our models for the lens systems and
a mass surface density $\sqrt{2}$ times that for a Salpeter mass
function.  We assign the observed fractional error in the less
magnified minimum image, taken from Table 4, to the simulated flux for
that miminum.  Since the less magnified minimum usually has the
narrowest magnification histogram, we expect the least variation in
its flux.  We use this to assign errors to the remaining simulated
fluxes assuming counting statistics.

We simulated each system 999 times.  For each of our 999 experiments we
compute the maximum likelihood value for the derived multiplicative
factor $\cal F$.  A histogram is shown in Figure 3.  The median
maximum likelihood value is 1.530, eight percent higher than the input
value.  The range of values including 68\% of the experiments runs
from 1.01 to 2.49, consistent with the width of the likelihood
function for the present data.
 
\subsection{Effects of individual systems on results}

We have examined the effects of the individual systems and find that one
system, RX J0911+0551, has a particularly strong influence on the
final result.  If we exclude it from the analysis the calibration
factor ${\cal F}$ is smaller by a factor of 1.5.

The pull of RX J0911+0551 on our principal result manifests itself
in many of the above tests.  For example, the effect of
a change in the power law index is much smaller when RX J0911+0551
is eliminated from the sample.

There are several arguments for excluding the system.  First,
the integrated likelihood for the system is more than a factor of ten
smaller than the system with the next smallest likelihood.   Second,
we carried out a second set of 999 simulations of RX J0911+0551
with an input calibration factor ${\cal F} = 1.189$.  In only 2.5\%
of the cases simulated was the likelihood of the input
value smaller than that computed using the observed X-ray flux ratios.
Third, at 150 days, the time delay between the $d$
image and the other three images is the longest for any of the systems
in the sample, rendering it more vulnerable to intrinsic X-ray variability.

The low likelihood for the present observations is explained by the
fact that the ratio observed for the flux of image $b$ to that of
image $d$ is 1.27, very much smaller than the 5.40 predicted by the
lens model.  A crude reduction of STIS spectra obtained as part of HST
program GO-9854 shows a C III] line strength ratio of 3.70, roughly
  consistent with the lens model, as would be expected if the broad
  line region is comparable to or larger in size than the Einstein
  rings of the individual stars.

While intrinsic X-ray variability is an attractive explanation, there
is nothing in the optical light curves presented by Hjorth et
al. (2002) to suggest a dramatic change in either image $b$ or image
$d$.  Our X-ray observations were obtained on JD 2451485.  If
intrinisic X-ray variation were to explain our unlikely result, then
image $d$ must have gotten substantially fainter in the previous 150
days, and $b$ must have gotten substantially fainter in the subsequent
150 days.  The Hjorth et al. (2002) light curves show them both getting
fainter by roughly 0.1 magnitude, very much less than would be needed
to explain our unlikely flux ratio.  The intrinsic variation at X-ray
wavelengths would need to be a factor of ten larger than at
optical wavelengths for this to have produced our result.

Image $d$ lies very far from the lensing galaxy and is a minimum of
the light travel time.  We therefore expect it to suffer very little
micro-lensing.  Image $b$ lies closer to the lensing galaxy, but as it
is also a minimum, the micro-lensing distribution function lacks the
tail toward fainter images that a similarly magnified saddle point
image would have.  The histogram is broadened, however, for higher stellar
densities, making those more likely and driving up the calibration
factor ${\cal F}$.

The above arguments constitute a rationalization for ignoring RX
J0911+0551 rather than a reason. 

 A toy model helps make the case for the inclusion
of  RX J0911+0551 in our result.  We suppose we have 10 systems, all of
  which have Gaussian likelihood distributions with the same standard
  deviation $\sigma$.  The uncertainty in the mean will be
  $\sigma/\sqrt{10 - 1}$, or $\sigma/3$.  The fact that 2.5\% of our
  simulated systems have likelihoods smaller than that of RX J0911+0511 would
  mean, under the Gaussian hypothesis, that RX J0911+0511 deviates from
  the true value by roughly $2.25\sigma$.  A single $2.25 \sigma$ point
  will pull the mean of ten objects $0.225 \sigma$, about 2/3
  of the uncertainty in the mean.  This is roughly  what RX J0911+0511
  does to the median likelihood for our ten systems.

\subsection{Cumulative effect of systematic errors}

In \S7.1 we give a statistical confidence interval for our result.  In
the subsequent subsections we consider a variety of systematic effects
that might be thought to affect our result.  Their cumulative effect
might be gauged by first making a best estimate of the range and
probability distribution of values within that range for the parameter
that controls each sytematic effect, and then choosing values for each
of parameter at random, reanalyzing the data, and accumulating a
likelihood distribution.

We have instead made crude guesses of a plausible value for each
parameter and carried out a single analysis with that parameter.  If
those rough guesses were correct, the largest sources of systematic
error, would be, in descending order: a systematic error in the exponent of our power
law model for the total surface mass density (32\%), a systematic error
in the uncertainties in our X-ray fluxes (24\%), and a systematic error
in our adopted measurements of effective radii for the lensing galaxies (23\%),
These are still relatively small compared with our statistical
uncertainty.  At such time as the number of systems analyzed increases
and the statistical uncertainty decreases, one will want to analyze these
systematic effects more carefully.  

We note that in several cases, in particular that of the power law
exponent, RX J0911+0551 plays the dominant role in determining the
size of the systematic effect.  One reasonably expect that with a larger
sample size these systematic errors might also decrease.

\section {Discussion} 

\subsection{The dark matter fraction of a typical galaxy} 

We have used the stellar mass fundamental plane to combine results
from the ten lensing galaxies in our sample, since the likelihood
distribution for the stellar mass surface density for any single
system is quite broad.  The single parameter that we constrain in this
process is a calibrating factor ${\cal F}$ for photometrically derived stellar
masses.

Previous investigators, ourselves included, have instead focused on
the dark matter fraction.  Our calibrated stellar mass fundamental
plane can be used to derive a projected dark matter fraction as a
function of $r/r_e$ for a specific choice of $\sigma_{prox}$ and
$r_e$, making the comparison with previous work more straightforward.

In constructing our fundamental plane we chose a fiducial galaxy with
$\sigma_{prox} = 266$ km/s and $r_e = 6.17$ kpc.
The stellar surface mass density at the effective radius, $\Sigma_e$
for such a galaxy is then $c{\cal F}/3.607$, or $5.04 \times 10^8 {\rm M_\odot/kpc^2}$ using the SL2S result for the constant $c$ in Table 1.

By assumption our galaxies have singular isothermal ellipsoidal mass
distributions with the same ellipticity as the observed light, so that
the stellar mass and dark matter fractions depend only upon the ratio
of the circularized radial coordinate $w$ defined by equation (10) to
the effective radius.  The projected dark matter fraction at the effective
radius for our fiducial galaxy is 62\%.

In Figure 4 we show the stellar mass fraction as a function of
$w/r_e$.  The stellar mass fraction reaches a maximum at
$w/r_e = 0.074$, the point at which the derivative of the de
Vaucouleurs surface brightness profile is equal to that of the
projected isothermal sphere.  The maximum stellar mass fraction
of 110\% indicates the breakdown of one or more of our model
assumptions: the isothermal profile for the combined mass, the
de Vaucouleurs profile for the stellar mass, or the invariance
with radius of the IMF (Mart\'in-Navarro et al. 2014).  But the breakdown is far
inside the effective radius, affecting of order 10\% of the
mass inside the Einstein radius, and ought not have a subtantial
effect on our principal result.

\subsection{Comparison with previous micro-lensing results}

The system Q2237+0305, ``Huchra's Lens,'' has been studied by Kochanek
(2004), who computes relative likelihoods for the stellar mass
contribution of the convergence on a coarse grid and finds it 
  five times as likely that $\kappa^*/\kappa $ is unity as opposed to
  0.5, implying a stellar fraction close to 100\%.

This system is quite unlike the systems considered here: with a lens
redshift of 0.04 the Einstein ring projects to a radius of only
several hundred parsecs on the galaxy, where one might expect most of
the mass to be in the form of stars.  Moreover the system is a barred
spiral, not an elliptical.  The Kochanek result, while broad, is
nonetheless consistent with our curve in Figure 4.

A second consequence of the low redshift of the lens is that the
Einstein rings of individual stars project to a larger scale on the
background quasar.  As a result even the optical emission region can
be treated as a point source.  And the velocities of stars within the
lens project to much larger velocities at the redshift of the quasar.
The timescale for variation is therefore much shorter than for the
systems in the present sample.

Using optical data, Bate et al. (2011) have determined projected
smooth (dark) matter fractions for the systems Q2237+0305, MG 0414+534
and SDSS J0924+0219.  The limits for the first two of these are quite
broad, but for SDSS J0924+0219 they obtain a projected smooth
(dark) matter fraction at the Einstein ring, $\approx 1.97 r_e$ of
$80 \pm 10$\%.  The dark matter fraction at this radius for our fiducial lens
is found from Figure 4 to be 82\%.  The prediction for a Salpeter IMF
stellar mass fundamental plane can be found by dividing the stellar
fraction in Figure 4 by our calibration factor ${\cal F} =
1.229$, yielding a dark matter fraction of 85\%.  The predictions for a
lens with a proxy dispersion of 214 km/s and an effective radius of
6.94 kpc are, by coincidence, identical to those for our fiducial
lens.  The Bate et al. (2011) results for SDSS J0924+0219 are therfore
consistent both with a Salpeter IMF and with our calibrated Salpeter
IMF.  

Where we use the stellar mass fundamental plane to predict the dark
matter fraction at the Einstein ring, Bate et al. (2011) use a different
scheme that they find underpredicts the smooth (dark matter) fraction.  They
start with the Bernardi et al. (2003) relation between surface
brightness and effective radius (the Kormendy relation) and convert
from light to mass using the Kauffmann et al. (2003) relation between
stellar mass to light ratio and absolute magnitude.  Kauffmann et
al. (2003) use a Kroupa stellar mass function.  A Salpeter IMF would have
predicted yet smaller smooth mass surface densities.  The fundamental
plane gives better predictions of surface brightness than the Kormendy
relation by virtue of using both effective radius {\it and} velocity
dispersion, so we should not be surprised that the dark matter fraction
predicted in \S 8.1 is closer to what Bate et al. (2011) observed.

Morgan et al. (2008) carry out a joint X-ray and optical analysis of
multi-epoch data for PG 1115+080.  They parameterize their result by
the stellar contribution to the lensing model.  For the maximum
likelihood value of their parameter, they find a stellar fraction at
the Einstein ring of 0.115, with an uncertainty of roughly a factor of
two in either direction.  Using our adopted value for the effective
radius, we find $\theta_{Ein}/\theta_e = 1.56$.  For the fiducial galaxy in
Figure 4 we find a stellar fraction of 0.23.  Using the measured
$r_e$ and $\sigma_{prox}$ we find a stellar fraction of 0.33.

But had we adopted the {\tt imfitfits} measurement of the effective
radius ($r_e= 4.22~{\rm kpc}$), which is what Morgan et al. (2008)
used, we would have $\theta_{Ein}/\theta_e = 2.19$, for which our fiducial
galaxy would give us a stellar fraction of 15\%.  Using the {\tt
  imfitfits} measured values $r_e= 4.22$ kpc and
$\sigma_{prox} = 232$ km/s we find a stellar fraction of 18\%,
which would appear to be consistent with the Morgan et al. (2008)
result.

Dai et al. (2010) carry out a joint X-ray and optical analysis of
multi-epoch data for RX J1131-1231.  They allow explicitly for errors
in the model magnifications.  Under their less conservative
priors they find a stellar contribution to the convergence of
30\% at 1.50 $r_e$.  This is consistent with the prediction of 26\% for
our fiducial galaxy.  Using instead the observed effective radius of 5.24 kpc
and a proxy dispersion of 349 km/s, we find again a 26\% prediction for the
stellar component. 

Mediavilla et al. (2009) have carried out an analysis of optical flux
ratios for 29 pairs of images of 20 lensed quasars.  When they allow
for a finite source size of $1.2 \times 10^{16}$ cm, consistent with
Pooley et al. (2007), the find (their Figure 8) a stellar mass
fraction of 0.10 with an uncertainty of roughly 0.05.  This is twice
as large as when they allow for a source only half as large.  For the
latter case, when they allow for errors of 0.2 mag in their magnitude
differences, they again find a stellar mass fraction of 0.10.  They do
not present results assuming both magnitude errors and the larger
source size.

Mediavilla et al. (2009) compare optical continuum flux ratios
to emission line ratios, mostly broad line.  Sluse et al. (2012b)
have argued that broad lines are micro-lensed, in which case
the use of broad-line flux ratios rather than model fluxes
will reduce the inferred effect of micro-lensing.

The majority of quasars in the Mediavilla sample are doubly imaged,
for which the images are more broadly scatterd to radii larger and
smaller than the Einstein radius.  They do not give effective radii,
but for $\theta_{Ein}/\theta_e \sim 1.5$, as is typical of our sample, we
would predict substantially larger stellar mass fractions of $\sim
0.25$.  

The present work includes several refinements over the Mediavilla et
al. (2009) effort: 1) we parameterize by mass-to-light ratio rather
than stellar fraction, accommodating a range in radii and elliptical
galaxies; 2) we use X-ray flux ratios, on the hypothesis that 
the X-ray emitting regions are point-like;
3) we use a substantially finer grid of magnification
simulations and interpolate among them.  The largest magnification in
the Mediavilla et al. (2009) simulations was 10, yet they have images
for which the magnification is greater than 20.  The next largest
magnification in their suite is 3.3.

Oguri, Rusu and Falco (2014) have carried out an analysis that has multiple
points in common with the present analysis.  They work with a sample of
strong lenses, including the lensed galaxies in the SLACS,
SL2S and BELLS (Brownstein et al. 2012) samples and also 28 of 
the lensed quasars in the CASTLES data base.
They compute stellar masses from the photometry and calculate
dynamical masses interior to the Einstein ring.  They add a power law
dark matter halo to bring the two into agreement.  They parameterize
the stellar contribution with a parameter $\alpha^{Sal}_{SPS}$,
that gives the ratio of the
stellar contribution to that computed from a Salpeter IMF, and which
is essentially our calibration factor $\cal F$.  They parameterize the
dark matter contribution by $A_{DM}$ which gives the ratio of the
projected dark matter contribution within the effective radius to the total
stellar mass.  These parameters are strongly degenerate.

Oguri et al. use the dark matter fraction likelihood distributions for
twelve quasars given by the present authors (Pooley et al. 2012) and three
given by Bate et al. (2011) to break this degeneracy.  They add an
additional parameter, $\gamma'_{DM}$, the slope of the density power
law adopted for the dark matter halo, which they also allow to vary.
They find $\alpha^{Sal}_{SPS} = 0.92^{+0.09}_{-0.08}$.

A particularly intriguing feature of this result is that the fractional
uncertainty in $\alpha^{Sal}_{SPS}$ is smaller by a factor of
five than the width of the likelihood for the stellar
fraction given in Pooley et al. (2012) and likewise that of the present
result for ${\cal F}$.  Figures 3 and 4 of Oguri et al. show a broad
range of allowed values for $\alpha^{Sal}_{SPS}$ in the absence of the
micro-lensing constraints.  An additional micro-lensing constraint
with an uncertainty of a factor of 1.5 in $\alpha^{Sal}_{SPS}$ would
not seem sufficient to narrow the allowed values to the claimed range
of only 10\%.

Oguri et al.  favor a power law exponent for the total mass density,
$\rho = \rho_0(r/r_0)^{-\gamma'}$, of $\gamma' = 2.11$, slightly
steeper than the isothermal assumed here.  As noted in \S7.7, using a
power law this steep would increase our calibration factor ${\cal F}$
by roughly 32\%, increasing the disagreement with the Oguri et al.
result.

Mediavilla et al. (2009), Bate et al. (2011) and Pooley et al. (2012)
all parameterized their results by a single dark matter fraction for
all four images.  In the present effort we have taken explicit account
of the differing distances of the quasar images from their host
galaxies, and of the ellipticity of the underlying light distribution.
The micro-lensing histograms are often quite different for images that
are saddle points and minima.  The minima tend to lie further from the
lensing galaxy, and therefore have lower stellar surface densities
than saddle points.  Moreover, to the extent that the ellipticity of
the light is aligned with the potential, the minima will tend to lie
along the stellar minor axis, producing yet lower stellar surface
densities.

The Oguri et al. result assumes a power law dark matter halo, where the
present result assumes a power law (isothermal) for the total surface
mass density.  In the present analysis, the power law index enters
primarily in the computed convergences and shears.  Changing that
index has an effect similar to adding or subtracting a mass sheet, as
described in \S7.6, driving the present result further from the Oguri
et al. result.  By the same token Pooley et al. (2012) used convergences
and shears appropriate to an isothermal.  Had they used $\gamma' =
2.11$, they would have deduced larger stellar mass fractions, driving
the Oguri et al. results closer to the present results.  

\section {Conclusions and Outlook}

We have determined the factor by which the stellar mass fundamental
plane at $z \sim 0.5$ must be increased if the anomalies in the
observed X-ray fluxes from our sample of quadruply macro-lensed
quasars are attributed to micro-lensing by the stars in the lensing
galaxy.

Constructing our stellar mass fundamental plane from the 
lensing early type galaxies in the SLACS and SL2S surveys,
using stellar masses as computed by Auger and collaborators
(Auger et al. 2010; Sonnenfeld et al. 2013) assuming a Salpeter
inital mass function, the factor by which the stellar mass fundamental
plane must be scaled is  ${\cal F} = 1.23$, with a one sigma confidence interval
$0.77 < {\cal F} < 2.10$

The result includes the mass due to white dwarfs, neutron stars,
stellar mass black holes, brown dwarfs and red dwarfs too faint to
contribute significantly to the observed light.

We have investigated a number of possible sources of systematic error
and estimated how large the effects might be.  

The greatest single source of uncertainty is sample size.  While this
can be addressed with the discovery and followup observation of
additional systems, to the extent that newly discovered systems are
not as bright as those in the present sample, they will be less
amenable to observations with Chandra.  Moreover the presumptive
follow-on mission to Chandra, Athena+, will not have the needed
angular resolution.

We note that for the systems in our sample that are faintest in the
optical, SDSS J0924+0219 and SDSS J1138+0314, the optical flux ratios
are nearly identical to the X-ray flux ratios.  This suggests that the
optical emission is coming from a region that is relatively small
compared to the Einstein rings of the stars, and that X-ray fluxes may
not be needed for this kind of analysis.

Beyond that there is a systematic uncertainty arising from systematic
differences in the effective radii measured for our lensing galaxies.
Observations with JWST might in principle resolve these, since the
point source PSF will be very much smaller than the image separations.
More extensive photometric observations with JWST would permit the
direct local calculation of the stellar surface mass density using an
assumed IMF, circumventing the stellar mass fundamental plane.

\acknowledgments We owe Professor Saul Rappaport a particular debt of
gratitude for raising the question of what one might hope to learn
from X-ray observations of lensed quasars.  P.L.S. and J.A.B
acknowledge support from NSF grants AST 02-06010 and AST 06-07601 and
Chandra grant GO7-8099.  P.L.S. gratefully acknowledges the
hospitality of Williams College during academic year 2013-14.  D.P.
gratefully acknowledges support from Chandra grants GO1-12135 and and
GO3-14102.  J.W. acknowledges with great pleasure generous support
received during his tenure as Schroedinger Visiting Professor 2013 at
the Pauli Center for Theoretical Studies which is supported by the
Swiss National Science Foundation (SNF), ETH Z\"urich and the
University of Z\"urich.

\clearpage
\begin{figure}
\includegraphics[angle=270,scale=0.65]{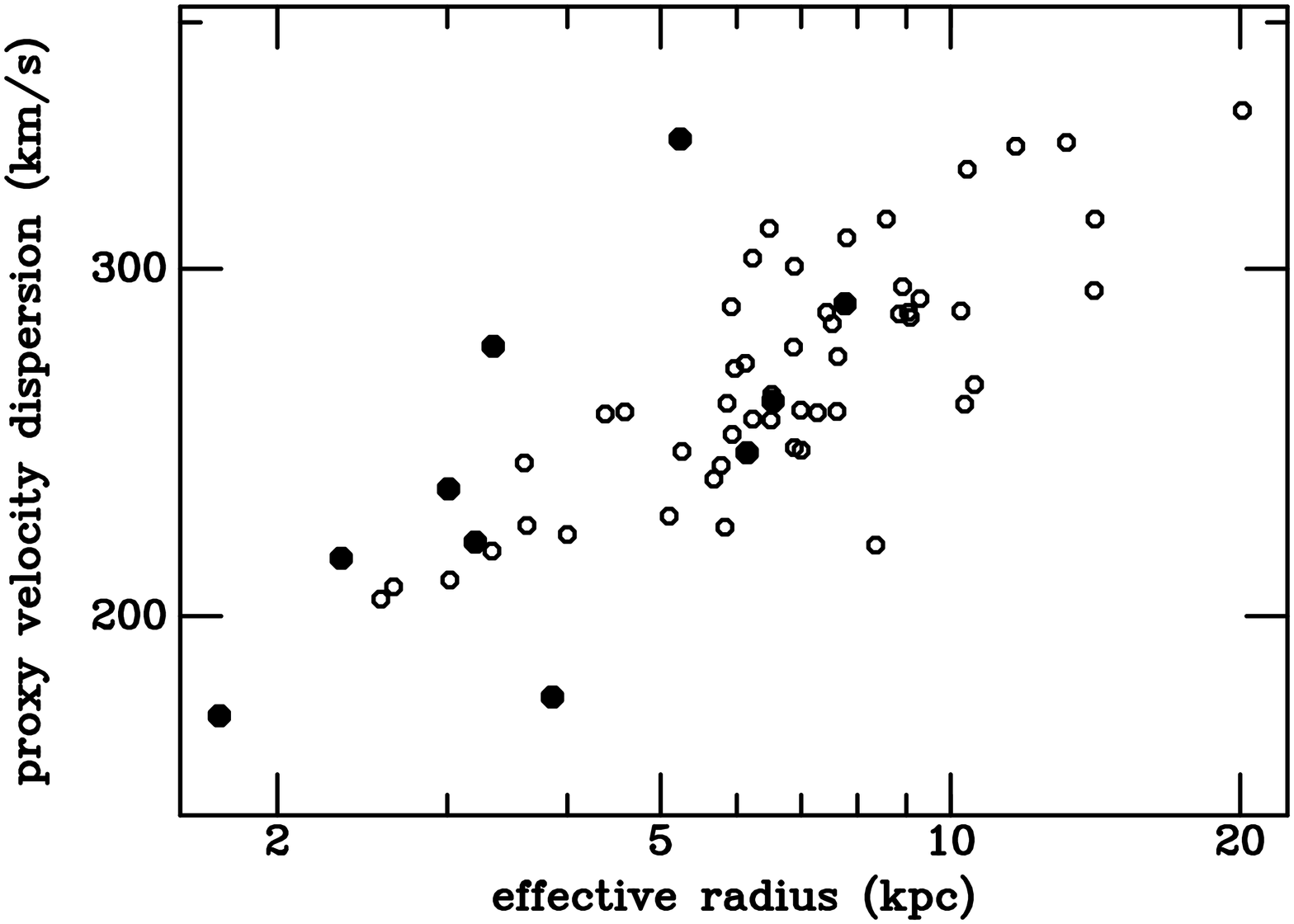}
\caption{Proxy velocity dispersions, computed from Einstein ring radii,
and photometric effective radii of lensing galaxies.  
The filled circles are for the
present sample, with proxy dispersions  and 
the adopted effective radii taken from Table 7.  The open circles are
from the SLACS sample. 
}
\end{figure}

\clearpage

\begin{figure}
\includegraphics[angle=270,scale=0.65] {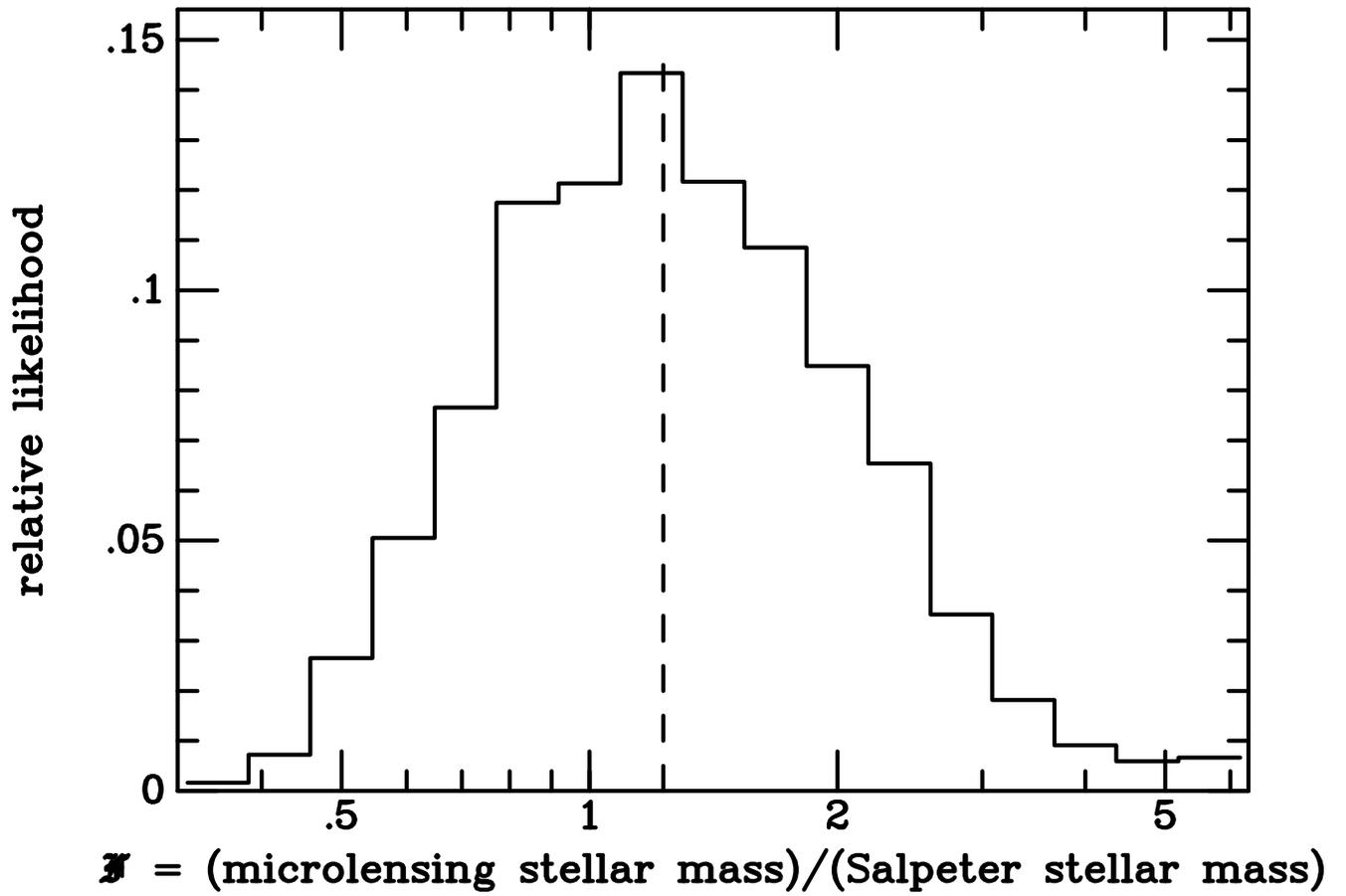}
\caption{Likelihood of the calibration factor $\cal{F}$ applied to
the stellar mass fundamental plane to compute 
the probability distribution of micro-lensing fluctations.
The dashed line shows the median likelihood value, ${\cal F} = 1.23$. 
}
\end{figure}

\clearpage

\begin{figure}
\includegraphics[angle=270, scale=0.65] {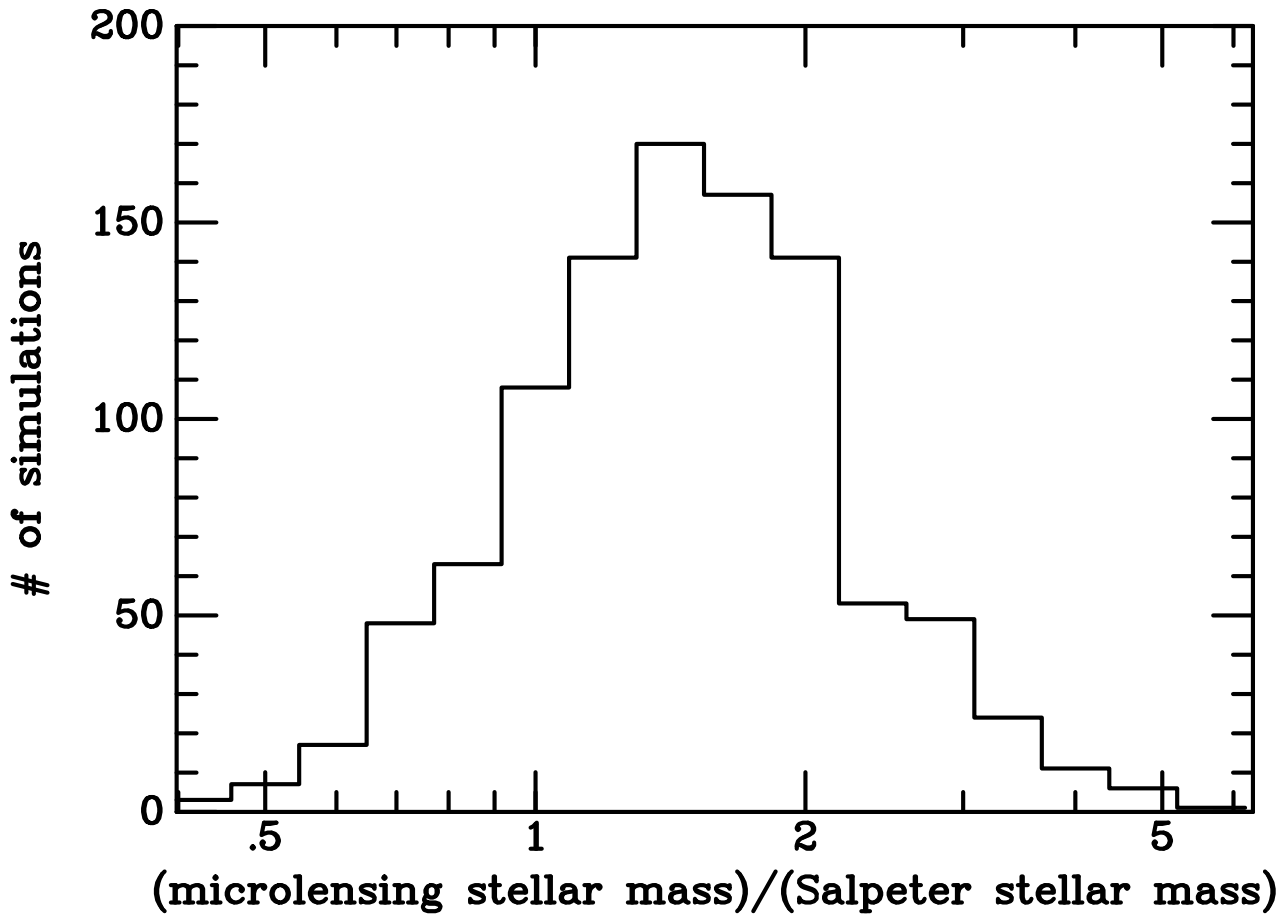}
\caption{Results of 999 simulations of our ten lensed systems,
adopting a value of 1.414 for the calibration factor $\cal{F}$ that multiplies
the stellar mass fundamental plane.
The bars give the numbers of simulations returning median 
likelihood values of $\cal{F}$ in the ranges shown.
}
\end{figure}

\clearpage

\begin{figure}
\includegraphics[angle=270, scale=0.65] {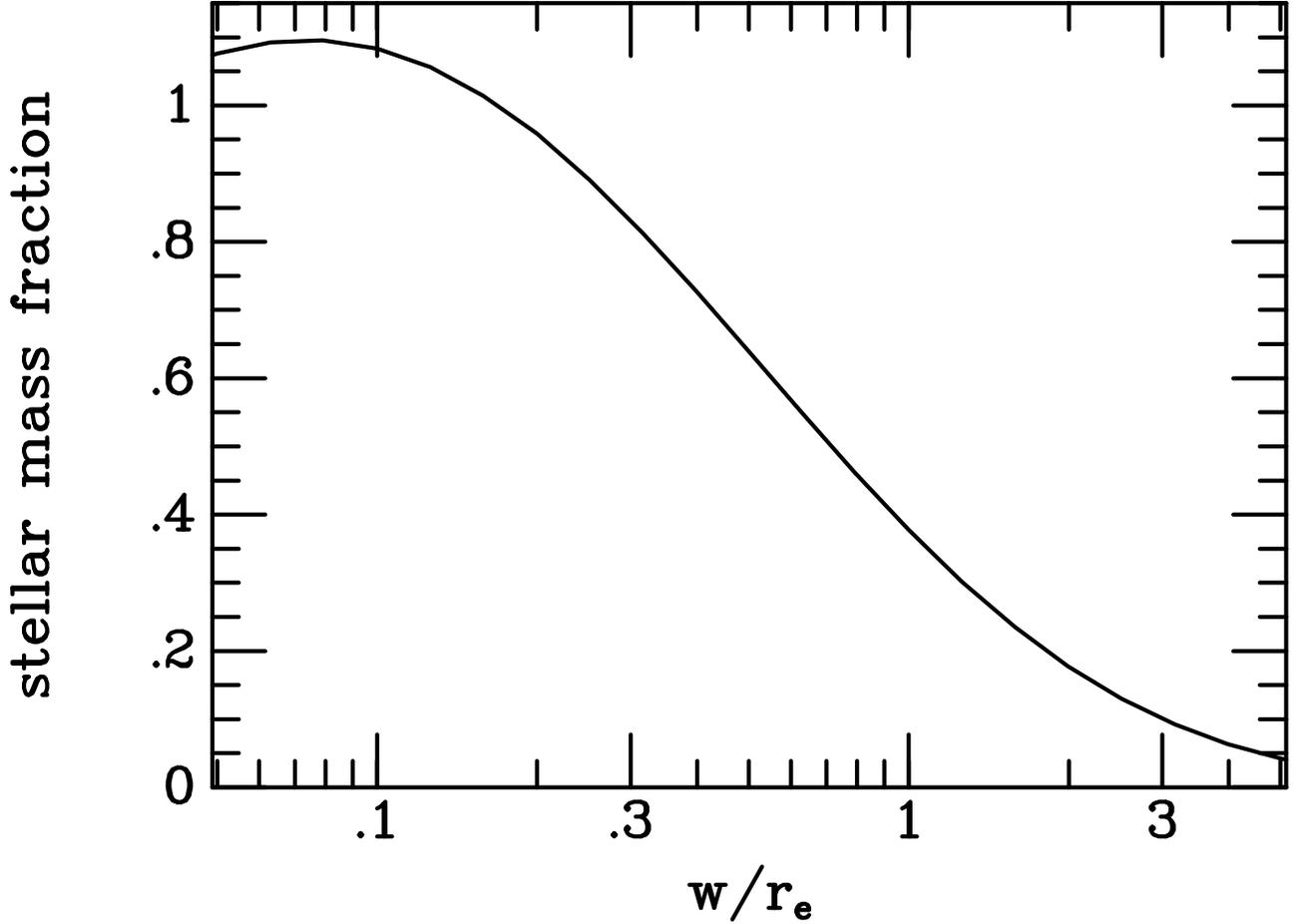}
\caption{The stellar surface mass density as a fraction of the
total for our fiducial galaxy, obtained by applying
our calibration factor $\cal{F}$ to the stellar mass
fundamental plane derived from the SLACS + SL2S samples.
The peak at $w/r_e = 0.074$ occurs where the slope of the  de Vaucouleurs
profile is equal to that of the singular isothermal.
}
\end{figure}
\clearpage
\begin{deluxetable}{lrrrrr}
\tablecolumns{6}
\tablewidth{0pc}
\tablecaption{Stellar Mass Fundamental Plane Coefficients}
\tablehead{
\multicolumn{6}{c}
{$\log \Sigma_e = a \log \left(\sigma_{prox} \over 266~{\rm km/s} \right)
+ b \log \left(r_e \over 6.17~{\rm kpc} \right) + c$}\\
\colhead{sample} & \colhead{a}   & \colhead{b}    & \colhead{c} &
\colhead{$r_{ab}$} & \colhead{rms}  \\
}
\startdata

SLACS$^*$             &  0.767     & -1.140     & 9.219 &    & 0.094 \\   
 & $\pm$0.219 & $\pm$0.086 & $\pm$ 0.014 & -0.622  &   \\

SLACS             &  1.590     & -1.371     & 9.216 &    & 0.076 \\   
 & $\pm$0.300 & $\pm$0.100 & $\pm$ 0.012 & -0.796  &   \\

SL2S              &  1.934     & -1.701     & 9.225        &  &  0.145 \\ 
 & $\pm$0.557 & $\pm$0.166 & $\pm$ 0.032 & -0.538 & \\

SLACS$^\dagger$   & $\equiv$1.748 & $\equiv$-1.453 & 9.220      &  & 0.077   \\
 & $\pm$0.260   & $\pm$0.085     & $\pm$0.011 & -0.758 &    \\

SL2S$^\dagger$  & $\equiv$1.748 & $\equiv$-1.453 & 9.170  &   & 0.154 \\   
 & $\pm$0.260   & $\pm$0.085     & $\pm$0.033 & -0.758 &    \\

\enddata
\tablenotetext{*}{Uses stellar velocity
dispersion $\sigma$  rather than $\sigma_{prox}$.}
\tablenotetext{\dagger}{$a$ and $b$ coefficients and errors are
values for combined SLACS + SL2S sample.}
\end{deluxetable}

\clearpage
\begin{deluxetable}{lrrrrr}
\tablecolumns{5}
\tablewidth{0pc}
\tablecaption{Stellar Mass Fundamental Line Coefficients}
\tablehead{
\multicolumn{5}{c}
{$\log \Sigma_e=A \log \left(\sigma_{prox} \over 265~{\rm km/s} \right) + C$}\\ 
\multicolumn{5}{c}
{$\log r_e=D \log \left(\sigma_{prox} \over 265~{\rm km/s} \right) + E$}\\
\colhead{sample} & \colhead{$A$}   & \colhead{C}    & \colhead{D} &
\colhead{E}  \\
}
\startdata
SLACS &  -1.702     & 9.156     & 2.401 &   0.832 \\   
    & $\pm$0.401 & $\pm$0.024 & $\pm$ 0.261 & 0.016    \\
\enddata
\end{deluxetable}

\clearpage
\begin{deluxetable}{llrllr}
\tablecolumns{6}
\tablewidth{0pc}
\tablecaption{X-ray Observations of Lensed Quasars}
\tablehead{
\colhead{object}  & \colhead{Date} & \colhead{ObsID} & 
\colhead{object}  & \colhead{Date} & \colhead{ObsID}  \\
}
\startdata
HE 0230-2130  & 2000 Oct 14 & 1642    & MG J0414+0534 & 2002 Jan 9& 3419  \\
HE 0435-1223  & 2006 Dec 17 & 7761    & RX J0911+0551 & 1999 Nov 3 & 419   \\
SDSS J0924+0219 & 2005 Feb 24 & 5604 & PG 1115+080&2000 Jun 3  & 363    \\
RX J1131-1231 &2004 Apr 12 & 4814  & SDSS J1138+0314& 2007 Feb 13 & 7759 \\
B1422+231 &2004 Dec 2    & 4939     & WFI J2033-4723  & 2005 Mar 10 &5603 \\  
\enddata
\end{deluxetable}

\clearpage
\begin{deluxetable}{llrrrrrrr}
\tablecolumns{9}
\tablewidth{0pc}
\tablecaption{Lens System Components: Relative Positions, Convergences, Shears,
Macromodel and Observed Relative X-ray Fluxes} 
\tablehead{
\multicolumn{2}{c}{object~~~component$^a$} &
\colhead{$x(\arcsec)$} & \colhead{$y(\arcsec)$} &
\colhead{$\kappa$} & \colhead{$\gamma$} &\colhead{$f_{mod}^b$}&
\colhead{$f_X^b$} & \colhead{$\sigma_{f_X}$}  \\
}
\startdata

HE 0230-2130
& g1 &          -0.072  & 1.085  & & & &  & \\
& g2 &          +0.212  & 2.059  & & & &  & \\
& a(HM)  &           0.000  & 0.000 & 0.472 & 0.416 &   1.91  & 1.58 & 0.2 \\
& b(HS)  &          -0.698  & 0.256 & 0.510 & 0.587 &   1.93  & 0.70 & 0.13\\
& c(LM)  &          -1.198  & 1.828 & 0.440 & 0.334 &   1.00  & 1.00 & 0.15\\
& d(LM)  &          +0.244  & 1.624 & 1.070 & 0.864 &   0.27  & 0.45 & 0.08\\

MG J0414+0534
& g  &          -0.472  &-1.277  & & & & & \\
& a1(HM) &           0.600  &-1.942 & 0.489 & 0.454 &   3.48  & 2.13 & 0.1 \\
& a2(HS) &           0.732  &-1.549 & 0.530 & 0.524 &   3.57  & 1.30 & 0.1 \\
& b(LM)  &           0.000  & 0.000 & 0.460 & 0.316 &   1.00  & 1.00 & 0.05\\
& c(LM)  &          -1.342  &-1.650 & 0.676 & 0.693 &   0.51  & 0.42 & 0.02\\

HE 0435-1223
& g  &          -1.165  &-0.573  & & & & & \\
& a(LM)  &           0.000  & 0.000 & 0.445 & 0.383 &   1.00  & 1.00 & 0.10\\
& b(HS)  &          -1.476  & 0.553 & 0.539 & 0.602 &   1.08  & .375 & .047\\
& c(HM)  &          -2.467  &-0.603 & 0.444 & 0.396 &   1.06  & .378 & .047\\
& d(LM)  &          -0.939  &-1.614 & 0.587 & 0.648 &   0.65  & .363 & .046\\

RX J0911+0551
& g  &          -0.698  & 0.512  & & & & & \\
& a(HS)  &           0.000  & 0.000 & 0.646 & 0.544 &   2.93  & 3.40 & 0.35\\
& b(HM)  &           0.260  & 0.406 & 0.586 & 0.281 &   5.41  & 1.27 & 0.04\\
& c(LM)  &          -0.018  & 0.960 & 0.637 & 0.577 &   2.49  & 0.35 & 0.12\\   
& d(LM)  &          -2.972  & 0.792 & 0.290 & 0.066 &   1.00  & 1.00 & 0.10\\    

SDSS J0924+0219
& g  &          -0.182  &-0.859  & & & & & \\
& a(HM)  &           0.000  & 0.000 & 0.472 & 0.456 &   2.31  & 3.15 & 0.7 \\
& b(LM)  &           0.061  &-1.805 & 0.443 & 0.383 &   1.00  & 1.00 & 0.25\\
& c(LM)  &          -0.968  &-0.676 & 0.570 & 0.591 &   0.99  & 0.42 & 0.17\\
& d(HS)  &           0.536  &-0.427 & 0.506 & 0.568 &   2.08  & 0.45 & 0.22\\

PG 1115+080
& g  &           0.381  &-1.344  & & & & & \\
& a1(HM) &           1.328  &-2.034 & 0.537 & 0.405 &   3.90  & 3.87 & 0.3 \\
& a2(HS) &           1.477  &-1.576 & 0.556 & 0.500 &   3.72  & 0.62 & 0.13\\
& b(LM)  &          -0.341  &-1.961 & 0.658 & 0.643 &   0.66  & 1.05 & 0.10\\
& c(LM)  &           0.000  & 0.000 & 0.472 & 0.287 &   1.00  & 1.00 & 0.10\\   

RX J1131-1231
& g  &          -1.444  & 1.706  & & & & \\
& a(HS)  &           0.588  & 1.120  & 0.494 & 0.562 &   1.73 & 0.22 & 0.025\\
& b(HM)  &           0.618  & 2.307 & 0.434 & 0.473  &   1.07  & 2.18 & 0.1 \\
& c(LM)  &           0.000  & 0.000 & 0.438 & 0.461  &   1.00  & 1.00 & 0.07\\
& d(LM)  &          -2.517  & 1.998  & 0.950 & 1.017 &   0.10 & 0.30 & 0.025\\

SDSS J1138+0314
& g  &           0.474  & 0.533  & & & & & \\
& a(HM)  &           0.     & 0.    & 0.465 & 0.384 &    1.40  & 3.20 & 1.0 \\
& b(LM)  &           0.103  & 0.979 & 0.578 & 0.673 &    0.71  & 1.00 & 0.4 \\
& c(LM)  &           1.184  & 0.812& 0.438 & 0.349 &     1.00   & 1.00 & 0.4\\
& d(HS)  &           0.698  &-0.056 & 0.523 & 0.614 &    1.30  & 1.30 & 0.5 \\ 

B1422+231
& g  &           0.742  &-0.656  & & & & & \\
& a(HM)  &           0.385  & 0.317 & 0.380 & 0.473 &    1.62  & 1.74 & 0.10\\
& b(HS)  &           0.000  & 0.000 & 0.492 & 0.628 &    1.91  & 0.95 & 0.08\\
& c(LM)  &          -0.336  &-0.750 & 0.365 & 0.378 &    1.00  & 1.00 & 0.10\\
& d(LM)  &           0.948  &-0.802 & 1.980 & 2.110 &    0.08  & 0.10 & 0.01\\

WFI J2033-4723
& g  &          -1.438  & 0.308  & & & & & \\
& a1(HM) &          -2.196  & 1.261 & 0.506 & 0.255 &  1.56  & 0.87 & 0.15\\
& a2(HS) &          -1.482  & 1.376 & 0.665 & 0.643 &  0.92  & 0.96 & 0.2 \\
& b(LM)  &           0.000  & 0.000 & 0.392 & 0.302 &  1.00  & 1.00 & 0.15\\
& c(LM)  &          -2.114  &-0.277 & 0.700 & 0.735 &  0.62  & 0.64 & 0.11\\

\enddata
\tablenotetext{a}{LM, HM, LS \& HS: the less magnified (L) and more highly magnified (H) minima (M) and saddle points (S) of the light travel time
surface.}
\tablenotetext{b}{Model and observed fluxes and errors are relative to the less magnified minimum, LM}
\end{deluxetable}

\clearpage
\begin{deluxetable}{lrrrrrr}
\tablecolumns{7}
\tablewidth{0pc}
\tablecaption{Photometric Properties of Lensing Galaxies}
\tablehead{
\colhead{object} & \colhead{$r_e(\arcsec)$} &  \colhead{$r_e(\arcsec)$} 
& \colhead{$r_e(\arcsec)$} & \colhead{$e$} & \colhead{PA$(^\circ)$} 
& \colhead{sources}\\
\multicolumn{2}{r}{({\tt imfitfits})} & \colhead{(MCS)} & \colhead{(adopted)} &
            & (E~of~N) & \colhead{for $r_e$} \\
}
\startdata
HE 0230-2130 G1  &      &      & 0.51  & 0.34  &  73.0 & a \\
HE 0230-2130 G2  &      &      & 0.65  & 0.24  & -46.0 & a \\
 MG J0414+0534    & 0.77 &      & 0.98  & 0.20  &  84.0 & b \\
 HE 0435-1223     & 0.86 & 1.50 & 1.13  & 0.09  &  -1.4 &  c, d \\
 RX J0911+0551    & 0.67 & 1.02 & 0.83  & 0.11  & -70.0 & b, e \\
SDSS J0924+0219  & 0.31 & 0.50 & 0.44  & 0.12  & -61.3 & l, f, g \\
 PG 1115+080      & 0.47 & 0.92 & 0.66  & 0.04  & -67.5 & b, e \\
 RX J1131-1231     &      & 1.51 & 1.19  & 0.25  & -71.4 & h  \\
 SDSS J1138+0314  &      & 0.86 & 0.67  & 0.16  & -57.3 & i \\
 B1422+231        & 0.31 & 0.41 & 0.36  & 0.39  & -58.9 & b    \\
 WFI J2033-4723   &      & 0.61 & 0.48  & 0.18  &  27.8 & j \\  
\enddata
\tablenotetext{a}{\ Levinson and Schechter, in preparation;
$^b$Kochanek et al (2000);
$^c$Kochanek et al (2006);
$^d$Courbin et al (2011);
$^e$Sluse et al (2012);
$^f$Eigenbrod et al (2006);
$^g$Keeton et al (2006);
$^h$Claeskens et al (2006; 
$^i$Chantry et al (2010); 
$^j$Vuissoz et al (2008); 
$^l$Morgan et al (2006)}
\end{deluxetable}

\clearpage

\begin{deluxetable}{lcccccccccccc}
\tablecolumns{13}
\tablewidth{430pt}
\tablecaption{Lens Model Parameters
  \label{tab:modelparams}}
\tablehead{
  \colhead{} &
  \multicolumn{3}{c}{Primary Lens} &
  \colhead{} &
  \multicolumn{2}{c}{Shear} &
  \colhead{} &
  \multicolumn{3}{c}{Secondary Lens} \\
  \cline{2-4}
  \cline{6-7}
  \cline{9-11}
  \cline{10-13}
  \colhead{Object} & 
  \colhead{$\theta_{Ein}$} & 
  \colhead{$e$} & 
  \colhead{$\phi_{e}$$^{a}$} & 
  \colhead{} &
  \colhead{$\gamma$} & 
  \colhead{$\phi_{\gamma}$$^{a}$} & 
  \colhead{} &
  \colhead{$b_2$} & 
  \colhead{$x_2$$^{c}$} & 
  \colhead{$y_2$$^{c}$} & 
}
\startdata
HE 0230-2130      & $0\farcs87$ & \nodata & \nodata && $0.112$                            & $-59\fdg9$           && $0\farcs33$ & $\phn\mbox{$-$}0.283$ & $+0.974$ \\ 
MG J0414+0534     & $1\farcs11$ & 0.20 & 84.0 && $0.063$                            & $+63\fdg6$           && $0\farcs14$ & $\phn\mbox{$-$}0.385$ & $+1.457$ \\ 
HE 0435-1223      & $1\farcs20$ & 0.09 &-1.40 && $0.063$                               & $-19\fdg8$           && \nodata     & \nodata               & \nodata  \\ 
RX J0911+0551     & $0\farcs95$ & 0.11 & -70.0 && $0.294$                            & $\phn\mbox{+}8\fdg3$ && $0\farcs22$ & $\phn\mbox{$-$}0.754$ & $+0.665$ \\ 
SDSS J0924+0219    & $0\farcs87$ & 0.12 & -61.3 && $0.064$                               & $+66\fdg2$           && \nodata     & \nodata               & \nodata  \\ 
PG 1115+080       & $1\farcs03$ & \nodata &\nodata && \nodata                               & \nodata              && $2\farcs56$ & $-10.883$             & $-5.266$ \\ 
RX J1131-1231     & $1\farcs78$ & 0.25 & -71.4 && $0.068$                            & $-77\fdg6$           && \nodata     & \nodata               & \nodata  \\ 
SDSS J1138+0314   & $0\farcs67$ & \nodata & \nodata && $0.098$                            & $+32\fdg6$           && \nodata     & \nodata               & \nodata  \\ 
B1422+231         & $0\farcs74$ & 0.39 & -58.9 && $0.137$ & $-47\fdg2$    && \nodata     & \nodata               & \nodata  \\ 
WFI J2033-4723    & $1\farcs06$ & 0.18 & 27.8 && $0.059$                            & $+45\fdg5$           && $0\farcs29$ & $\phn\mbox{$+$}0.229$ & $+2.020$ \\ 
\enddata
\tablenotetext{a}{Position angles of ellipticity, 
$\phi_e$ and external shear $\phi_\gamma$, measured
in degrees east of north.}
\tablenotetext{c}{Fixed position of secondary galaxy, relative to main
  lensing galaxy, in arcseconds. Allowed to vary azimuthally in the case
  of PG 1115+080; see text.  Consistent with {\tt lensmodel} and in
  contrast to the entries in Table 4, the positive directions of $x$ and $y$
  are west and north, respectively.}
\end{deluxetable}

\clearpage
\begin{deluxetable}{lcccc}
\tablecolumns{5}
\tablewidth{0pc}
\tablecaption{Redshifts, Effective Radii and Proxy Dispersions}
\tablehead{
\colhead{object} & \colhead{$z_l$} & \colhead{$z_s$} & \colhead{$r_e$} &
\colhead{$\sigma_{prox}$} \\
 & & & \colhead{(kpc)} & \colhead{(km/s)} \\
} 
\startdata
 HE 0230-2130 g1 & 0.523  & 2.163 & 3.21 & 218 \\
 ~~~~~~~~~~~~~~~~~~~g2 &        &       & 4.09 & 134 \\
 MG J0414+0534   & 0.96   & 2.64  & 7.77& 288 \\
 HE 0435-1223    & 0.4541 & 1.689 & 6.54 & 257 \\
 RX J0911+0551   & 0.77   & 2.80  & 6.15 & 242 \\
 SDSS J0924+0219 & 0.39   & 1.524 & 2.33 & 214 \\
 PG 1115+080     & 0.31   & 1.72  & 3.01 & 232 \\
 RX J1131-1231   & 0.295  & 0.658 & 5.24 & 349 \\
 SDSS J1138+0314 & 0.45   & 2.44  & 3.86 & 182 \\
 B1422+231       & 0.34   & 3.62  & 1.74 & 178 \\
 WFI J2033-4723  & 0.66   & 1.66  & 3.35 & 274 \\  
\enddata
\end{deluxetable}

\clearpage
\begin{deluxetable}{lrrrrr}
\tablecolumns{6}
\tablewidth{0pc}
\tablecaption{Deviations of Radio, Mid-IR and [O III] Fluxes from
Macro-model Predictions}
\tablehead{
\colhead{component} & \colhead{$f_X$} & \colhead{$f_{mod}$} 
& \colhead{$f_{mil}$}
& \colhead{$-2.5\log\left(f_{mil}\over f_{mod}\right)$}
& \colhead{$-2.5\log\left(f_X\over f_{mod}\right)$} }
\startdata
&\multicolumn{5}{l}{MG 0414+0534 (11.7 $\mu$; MacLeod et al. 2011)} \\
HM(a1)/LM &  2.13 &  3.56  &  2.96 &  +0.20 &  -0.56 \\
HS(a2)/LM &  1.30 &  3.88  &  2.74 &  +0.37 &  -1.18 \\
LS(c)/LM  &  0.42 &  0.32  &  0.43 &  -0.36 &   +0.33 \\
LM(b)/LM  & $\equiv 1$ & $\equiv 1$ & $\equiv 1$ & $\equiv 0$ &  $\equiv 0$  \\
&\multicolumn{5}{l}{PG 1115+080 (11.7 $\mu$; Chiba et al 2005)} \\
HM(a1)/LM &  3.87 &  3.75  & 4.76 & -0.26 &   +0.04 \\
HS(a2)/LM &  0.62 &  3.45  & 4.43 & -0.27 &   -1.86 \\
LS(b)/LM  &  1.05 &  0.78  & 0.76 & +0.03 &   +0.32 \\
LM(c)/LM  & $\equiv 1$ & $\equiv 1$ & $\equiv 1$ & $\equiv 0$ &  $\equiv 0$  \\
&\multicolumn{5}{l}{RX J1131-1231 ([O III]; Sugai et al 2007)} \\
HS(a)/LM  &  0.22 &  1.77  & 1.37 & +0.28 &  -2.26  \\
HM(b)/LM  &  2.18 &  1.04  & 0.84 & +0.23 &  +0.80  \\
LS(d)/LM  &  0.30 &  0.12  &  ... &  ...  &  +1.00  \\
LM(c)/LM  & $\equiv 1$ & $\equiv 1$ & $\equiv 1$ & $\equiv 0$ &  $\equiv 0$  \\
&\multicolumn{5}{l}{B1422+231 (8.4 GHz; Patnaik 1999)} \\
HM(a)/LM  & 1.74  &  1.56  & 1.88 & -0.20 & +0.12 \\
HS(b)/LM  & 0.95  &  2.12  & 2.02 & +0.05 & -0.87 \\
LS(d)/LM  & 0.10  &  0.06  & 0.06 &  0.00 & +0.52 \\
LM(c)/LM  & $\equiv 1$ & $\equiv 1$ & $\equiv 1$ & $\equiv 0$ &  $\equiv 0$  \\
\enddata
\end{deluxetable}

\end{document}